\documentclass[12pt]{article}
\usepackage{latexsym,cite,amssymb,amsthm,wasysym,textcomp,stmaryrd}
\usepackage{times}
\input amssym.def
\input amssym.tex

\makeatletter

\@addtoreset{equation}{section} \makeatother

\def\ta{\tilde{{\rm a}}}
\def\a{{\rm a}}

\def\dis{{\displaystyle}}

\def\B{{\cal B}}

\def\E{{\cal E}}

\def\C{{\cal C}}
\def\ta{\tilde{{\rm a}}}
\def\c{{\rm c}}
\def\tc{\tilde{{\rm c}}}

\def\dalpha{{\dot{\alpha}}}
\def\dbeta{{\dot{\beta}}}
\def\dgamma{{\dot{\gamma}}}
\def\ddelta{{\dot{\delta}}}

\def\diag{{\rm diag}}

\def\N{{\cal N}}

\def\half{{{\textstyle\frac{1}{2}}}}

\newcommand{\be}{\begin{equation}}
\newcommand{\ee}{\end{equation}}
\newcommand{\ba}{\begin{array}}
\newcommand{\ea}{\end{array}}

\newcommand{\su}{\mbox{su}}
\newcommand{\so}{\mbox{so}}

\def\A{{\cal A}}
\def\B{{\cal B}}
\def\C{{\cal C}}

\def\AA{\bf{A}}

\def\CC{\bf{C}}

\def\E{{\cal E}}
\def\L{{\cal L}}

\def\Q{{\cal Q}}

\def\DP{\Delta_{\Psi}}

\def\VE{\xi}

\def\tr{{\rm tr}}

\def\I_N{{1_{\scriptscriptstyle N\times N}}}

\def\5D{{\scriptscriptstyle 5D}}
\def\N{{\cal  N}}

\def\Z{\hat{Z}}

\def\I_N{{1_{\scriptscriptstyle N\times N}}}

\def\diag{{\rm diag}}

\begin{document}
\begin{titlepage}
\title{\vskip -60pt
{\small
\begin{flushright}
%hep-th/yymmnnn
\end{flushright}}
\vskip 30pt
{\textbf{Lecture note on Clifford algebra}}
\vskip
10pt
\author{Jeong-Hyuck Park}}
\date{}
\maketitle \vspace{-1.0cm}
\begin{center}
~~~\\
Department of Physics, Sogang University, 35 Baekbeom-ro, Mapo-gu, Seoul 04107,  Korea\\
~{}\\
{{\small{{{park@sogang.ac.kr}}}
{{{}}}}}\\
~~~\\
~~~\\
\end{center}
\begin{abstract}
\noindent This lecture note surveys the gamma matrices in general dimensions with arbitrary signatures, the study  of which is essential to understand   the  supersymmetry in the corresponding spacetime. The contents supplement the lecture presented by the author at
\textit{Modave Summer School in Mathematical Physics, Belgium}, june, 2005.
\end{abstract}
\today

\vspace{1cm}
%%\keywords{gamma matrix, supersymmetry, octonion}

\end{titlepage}
\newpage
%\tableofcontents
%\newpage

\section{Preliminary}
{\bf{Where do we see Clifford algebra?}}
\begin{itemize}
\item
Dirac equation,~  for sure.
\item
Supersymmetry algebra.
\item
Non-anti-commutative superspace.
\item
Division algebra, ${\mathbb{R,C,H, O}}$.
\item
Atiyah--Drinfeld--Hitchin--Manin  construction of Yang--Mills instantons, $F=\pm\ast F$.
\end{itemize}
~\\
The gamma matrices in  the Euclidean two-dimensions provide the \textit{fermionic oscillators},
\be
\ba{lll}
f^{2}=0\,,~~~&~~~\bar{f}{}^{\,2}=0\,,~~~&~~~\{f,\bar{f}\}=1\,,
\ea
\ee
where $f=\half(\gamma^{1}+i\gamma^{2})$, $\bar{f}=\half(\gamma^{1}-i\gamma^{2})$. Consequently, the irreducible representation is given uniquely by $2\times 2$ matrices acting on two dimensional spinors, $|+\rangle$ and $|-\rangle$,
\be
\ba{ll}
f=|-\rangle\langle +|=\left(\ba{ll}0&0\\1&0\ea\right)\,,~~~~&~~~~\bar{f}=|+\rangle\langle -|=\left(\ba{ll}0&1\\0&0\ea\right)\,.
\ea
\ee
Higher dimensional gamma matrices are then constructed by the direct products of them.

\section{Gamma Matrix}
We start with the following  \textbf{Theorem} on linear
algebra.\newline
\newline
\textbf{Theorem}\newline Any matrix,~$M$, satisfying
$M^{2}=\lambda^{2}\neq 0$, $\lambda\in {\mathbf C} $ is
diagonalizable, and furthermore if  there is another invertible
matrix, $N$, which anti-commutes with $M$, $\{N,M\}=0$, then $M$ is
$2n\times 2n$  matrix of the form
\begin{equation}
M=S\left(\begin{array}{cc}
         \lambda&0\\
         0&-\lambda
          \end{array}\right)S^{-1}\,.
\label{lemma}
\end{equation}
In particular, $\mbox{tr}M=0$. See Sec.\,\ref{Proof} for our proof.

%%%%%%%%%%%%%%%%%%%%%%%%%%%%%%%%%%%%%%%%%%%%%%%%%%%%%%%%%%%%%%%%%%%%%%%%%%%%%

%%%%%%%%%%%%%%%%%%%%%%%
%%%%%%%%%%%%%%%%%%%%%%%%%%%%%%%%%%%%%%%%%%%%
%%%%%%%%%%%%%%%%%%%%%%%%%%%%%%%%%%%%%%%%%%%%%%
\subsection{In Even Dimensions}
In even $d=t+s$ dimensions, with metric\footnote{Note that
throughout the lecture note we adopt the field theorists' convention
rather than string theorists such that the time directions have the
positive signature. The conversion is straightforward.}
\begin{equation}
\eta^{\mu\nu}=\mbox{diag}(\underbrace{++\cdots +}_{{\displaystyle t}}
\underbrace{--\cdots -}_{{\displaystyle s}})\,,
\end{equation}
gamma matrices,~$\gamma^{\mu}$, satisfy the Clifford algebra
\begin{equation}
\gamma^{\mu}\gamma^{\nu}+\gamma^{\nu}\gamma^{\mu}=2\eta^{\mu\nu}\,.
\end{equation}
With\footnote{``[\,]'' means  the standard anti-symmetrization with
``strength one''.}
\begin{equation}
\gamma^{\mu_{1}\mu_{2}\cdots\mu_{m}}=\gamma^{[\mu_{1}}\gamma^{\mu_{2}}\cdots
\gamma^{\mu_{m}]}\,,
\end{equation}
we define $\Gamma^{M},\,M=1,2,\cdots 2^{d}$
by assigning numbers to independent
$\gamma^{\mu_{1}\mu_{2}\cdots\mu_{m}}$, \textit{e.g.~}imposing
$\mu_{1}<\mu_{2}<\cdots <\mu_{m}$,
\begin{equation}
\Gamma^{M}=(1,\gamma^{\mu},\gamma^{\mu\nu},\cdots,\gamma^{\mu_{1}\mu_{2}\cdots
\mu_{m}},\cdots,\gamma^{12\cdots d})\,.
\end{equation}
Then $\{\Gamma^{M}\}/Z_{2}$ forms a group:
\begin{equation}
\begin{array}{cc}
\Gamma^{M}\Gamma^{N}=\Omega^{MN}\Gamma^{L}\,,~~~&~~~\Omega^{MN}=\pm 1\,,
\end{array}
\label{GGK}
\end{equation}
where $L$ is a fuction of $M,N$ and $\Omega_{MN}=\pm 1$ does not
depend on the specific choice of representation of the gamma
matrices. \newline \textbf{Theorem}~(\ref{lemma}) implies
\begin{equation}
\frac{1}{2n}\mbox{tr}(\Gamma^{M}\Gamma^{N})=\Omega_{MN}\delta^{MN}\,,
\end{equation}
which shows the linear independence of $\{\Gamma^{M}\}$ so that  any gamma
matrix should not be smaller than $2^{d/2}\times
2^{d/2}$. \newline

In two-dimensions, one can take the Pauli sigma matrices,
$\sigma^{1},\sigma^{2}$ as gamma matrices
with a possible factor,~$i$, depending on the signature.    In general,
 one can construct  $d+2$ dimensional gamma
matrices from $d$ dimensional gamma matrices by taking tensor products as
\begin{equation}
(\gamma^{\mu}\otimes\sigma^{1},~1\otimes\sigma^{2},~1\otimes\sigma^{3})
~~~~~~~\mbox{:~up~to~a~factor~}i\,.
\label{dd+2}
\end{equation}
Thus, the smallest size of irreducible representations
is  $2^{d/2}\times 2^{d/2}$
and $\{\Gamma^{M}\}$ forms a basis of $2^{d/2}\times
2^{d/2}$  matrices. \newline

By  induction on the dimensions, from eq.(\ref{dd+2}),  we may require gamma matrices to satisfy the
hermiticity condition
\begin{equation}
\gamma^{\mu}{}^{\dagger}=\gamma_{\mu}=
\left\{\ba{ll}+\gamma^{\mu}~~&~~\mbox{for~time-like~}\mu\\
{}&{}\\
-\gamma^{\mu}~~&~~\mbox{for~space-like~}\mu\ea\right.\,.
\end{equation}
With this choice of gamma matrices we define $\gamma^{(d+1)}$ as
\begin{equation}
\gamma^{(d+1)}=\sqrt{(-1)^{\frac{t-
s}{2}}}\gamma^{1}\gamma^{2}\cdots\gamma^{d}\,,
\end{equation}
satisfying
\begin{equation}
\begin{array}{c}
\gamma^{(d+1)}=(\gamma^{(d+1)})^{-1}=\gamma^{(d+1)}{}^{\dagger}\,,\\
{}\\
\{\gamma^{\mu},\gamma^{(d+1)}\}=0\,.
\end{array}
\label{gammad+1}
\end{equation}

For two sets of  irreducible gamma matrices,
$\gamma^{\mu},\,\gamma^{\prime}{}^{\mu}$  which are $2n\times 2n,\,
2n^{\prime}\times 2n^{\prime}$ respectively,  we consider a matrix
\begin{equation}
S=\sum_{M}\Gamma^{\prime}{}^{M}T(\Gamma^{M})^{-1}\,,
\label{Sform}
\end{equation}
where $T$, is an arbitrary $2n^{\prime}\times 2n$ matrix. \newline
This matrix satisfies for any $N$ from eq.(\ref{GGK})
\begin{equation}
\Gamma^{\prime}{}^{N}S=S\Gamma^{N}\,.
\label{similarity}
\end{equation}
By Schur's  Lemmas, it should be either $S=0$ or $n=n^{\prime}, \det{S}\neq
0$. Furthermore, $S$ is unique up to constant, although $T$ is
arbitrary.  This
implies the  uniqueness of the irreducible $2^{d/2}\times
2^{d/2}$ gamma
matrices in even $d$ dimensions, up to the similarity
transformations. These similarity transformations are also unique up to
constant. Consequently there exist similarity transformations which  relate
$\gamma^{\mu}$ to $\gamma^{\mu}{}^{\dagger},\,\gamma^{\mu}{}^{\ast},\,
\gamma^{\mu}{}^{T}$ since the latter  form
also representations of the Clifford algebra. By combining
$\gamma^{(d+1)}$ with  the similarity transformations, from
eq.(\ref{gammad+1}), we may acquire  the opposite sign,
$-\gamma^{\mu}{}^{\dagger},\,-\gamma^{\mu}{}^{\ast},\,
-\gamma^{\mu}{}^{T}$ as well. \newline
%%%and  hence there exist
%%%$A_{\pm},B_{\pm},C_{\pm}$ such that
%%%\begin{eqnarray}
%%%\pm\gamma^{\mu}{}^{\dagger}=A_{\pm}\gamma^{\mu}A_{\pm}^{-1}\label{A}\\
%%%{}\nonumber\\
%%%\pm\gamma^{\mu}{}^{\ast}=B_{\pm}\gamma^{\mu}B_{\pm}^{-1}\label{B}\\
%%%{}\nonumber\\
%%%\pm\gamma^{\mu}{}^{T}=C_{\pm}\gamma^{\mu}C_{\pm}^{-1}\label{C}
%%%\end{eqnarray}
%%%$A_{\pm},B_{\pm},C_{\pm}$ can be chosen to be unitary  only if we
%%%rescale them properly
%%%\begin{equation}
%%%%\begin{array}{ccc}
%%%A_{\pm}^{\dagger}A_{\pm}=1~~~~&~~~
%%%%B_{\pm}^{\dagger}B_{\pm}=1~~~~&~~~C_{\pm}^{\dagger}C_{\pm}=1
%%%%\end{array}
%%%%\label{ABC}
%%%%\end{equation}
Explicitly
%%%\cite{kugotownsend}
we define\footnote{Alternatively, one
can construct $C_{\pm}$ explicitly out of the gamma matrices in a
certain representation~\cite{strathdee}.}
\begin{equation}
A=\sqrt{(-1)^{\frac{t(t-1)}{2}}}\gamma^{1}\gamma^{2}\cdots\gamma^{t}\,,
\label{defA}
\end{equation}
satisfying
\begin{eqnarray}
A=A^{-1}=A^{\dagger}\,,\label{AAA}\\
{}\nonumber\\
\gamma^{\mu}{}^{\dagger}=(-1)^{t+1}A\gamma^{\mu}A^{-1}\,.\label{AgA}
\end{eqnarray}

If we write
\begin{equation}
\pm\gamma^{\mu}{}^{\ast}=B_{\pm}\gamma^{\mu}B_{\pm}^{-1}\,,
\label{B}
\end{equation}
then from
\begin{equation}
\gamma^{\mu}=(\gamma^{\mu}{}^{\ast})^{\ast}=B_{\pm}^{\ast}B_{\pm}\gamma^{\mu}
(B_{\pm}^{\ast}B_{\pm})^{-1}\,,
\end{equation}
one can normalize $B_{\pm}$ to satisfy~\cite{kugotownsend,scherk}
\begin{eqnarray}
&B_{\pm}^{\ast}B_{\pm}=\varepsilon_{\pm}\,1\,,~~~~~~~~
\varepsilon_{\pm}=(-1)^{\frac{1}{8}(s-t)(s-t\pm 2)}\,,&\label{B*B}\\
&{}&\nonumber\\
&B^{\dagger}_{\pm}B_{\pm}=1\,,&\label{Buni}\\
&{}&\nonumber\\
&B_{\pm}^{T}=\varepsilon_{\pm}\,B_{\pm}\,,&
\end{eqnarray}
where the  unitarity   follows from
\begin{equation}
\gamma^{\mu}=\gamma_{\mu}^{\dagger}=(\pm B^{-
1}_{\pm}\gamma_{\mu}^{\ast}B_{\pm})^{\dagger}=\pm
B^{\dagger}_{\pm}\gamma^{\mu}{}^{\ast}(B^{\dagger}_{\pm})^{-1}
=B^{\dagger}_{\pm}B_{\pm}\gamma^{\mu}(B^{\dagger}_{\pm}B_{\pm})^{-1}\,,
\end{equation}
and the positive definiteness of $B^{\dagger}_{\pm}B_{\pm}$. The
calculation of $\varepsilon_{\pm}$ is essentially counting the
dimensions of symmetric and anti-symmetric
matrices~\cite{kugotownsend,scherk}\footnote{From (\ref{CgC}) we
have $(C_{\pm}\gamma^{\mu_{1}\mu_{2}\cdots\mu_{n}})^{T}=\chi_{n\pm
}\,C_{\pm} \gamma^{\mu_{1}\mu_{2}\cdots\mu_{n}}$, $\chi_{n\pm
}\!\!:=\varepsilon_{\pm} (\pm
1)^{t+n}(-1)^{n+\frac{1}{2}(t+n)(t+n-1)}$ (\ref{transpose}). Thus,
one can obtain the dimension of the symmetric $2^{d/2}\times
2^{d/2}$ matrices as
\[\displaystyle{
2^{d/2-1}\left(2^{d/2}+1\right)=\sum_{n=0}^{d}\half\left(1+\chi_{n\pm}\right)
\frac{d!}{\,n!(d-n)!\,}\,.}\] From this one can obtain the value of
$\varepsilon_{\pm}$ (\ref{B*B}).}.

What is worthy of notice  is the case $\varepsilon_{\pm}=+1$. As we see
later in (\ref{Reta1}), (\ref{Reta2}), \textbf{if
$\varepsilon_{+}=+1$, the gamma matrices can be chosen to real, i.e.
$B_{+}=1$, while if $\varepsilon_{-}=+1$, the gamma matrices can be
chosen to pure imaginary, i.e. $B_{-}=1$.} Especially when the gamma
matrices are real we say they are in the Majorana
representation.\newline

The charge conjugation matrix,\,$C_{\pm}$,  given by
\begin{equation}
C_{\pm}=B^{T}_{\pm}A\,,
\end{equation}
satisfies\footnote{Essentially
all the properties of the charge conjugation matrix,\,$C_{\pm}$
depends only on $d$ and $\zeta$. However it is useful here to have
expression in terms of the  signature  to dicuss the Majorana
supersymmetry  later.} from the properties of $A$ and
$B_{\pm}$
\begin{eqnarray}
~~~~~~~~~~C_{\pm}\gamma^{\mu}C_{\pm}^{-1}=\zeta\gamma^{\mu}{}^{T}\,,~~~~~~~~~~~~~~~
\zeta=\pm(-1)^{t+1}\,,
~~~~~~~~~\label{CgC}\\
{}\nonumber \\
C_{\pm}^{\dagger}C_{\pm}=1\,,~~~~~~~~~~~~~~~~~~~~~~~~~~~~~~~~\\
{}\nonumber\\
C^{T}_{\pm}=(-1)^{\frac{1}{8}d(d-\zeta 2)}\,C_{\pm}
=\varepsilon_{\pm}(\pm 1)^{t}(-1)^{\frac{1}{2}t(t-1)}\,C_{\pm}\,,
~~~~~~~\label{Ctranspose}\\
{}\nonumber\\
\zeta^{t}(-1)^{\frac{1}{2}t(t-1)}A^{T}=B_{\pm}AB_{\pm}^{-1}=
C_{\pm}AC_{\pm}^{-1}\,.~~~~~~~~~~~
\label{ABCpm}
\end{eqnarray}
The sign factors $\varepsilon_{\pm}$  and $\zeta$ are  related by
\begin{equation}
\varepsilon_{\pm}=\zeta^{t}(-1)^{\frac{1}{2}t(t-1)+\frac{1}{8}d(d-\zeta
2)}\,.
\end{equation}
Eqs.(\ref{CgC},\,\ref{Ctranspose}) imply
\begin{equation}
\begin{array}{ll}
(C_{\pm}\gamma^{\mu_{1}\mu_{2}\cdots\mu_{n}})^{T}
&=\zeta^{n}(-1)^{\frac{1}{8}d(d-\zeta 2)+\frac{1}{2}n(n-1)}
\,C_{\pm}\gamma^{\mu_{1}\mu_{2}\cdots\mu_{n}}\\
{}&{}\\
{}&=\varepsilon_{\pm}
(\pm 1)^{t+n}(-1)^{n+\frac{1}{2}(t+n)(t+n-1)}\,C_{\pm}
\gamma^{\mu_{1}\mu_{2}\cdots\mu_{n}}\,.
\end{array}
\label{transpose}
\end{equation}

$\gamma^{(d+1)}$ satisfies
\begin{equation}
\begin{array}{l}
\gamma^{(d+1)}{}^{\dagger}=(-
1)^{t}A_{\pm}\gamma^{(d+1)}A_{\pm}^{-1}=\gamma^{(d+1)}\,,\\
{}\\
\gamma^{(d+1)}{}^{\ast}=(-1)^{\frac{t-s}{2}}B_{\pm}\gamma^{(d+1)}B_{\pm}^{-1}\,,\\
{}\\
\gamma^{(d+1)}{}^{T}=(-1)^{\frac{t+s}{2}}C_{\pm}\gamma^{(d+1)}C_{\pm}^{-1}\,,
\end{array}
\label{daggerA}
\end{equation}
where $\{A_{+},A_{-}\}=\{A,\gamma^{(d+1)}A\}$.\newline

In stead of eq.(\ref{dd+2}) one can   construct  $d+2$ dimensional gamma
matrices from $d$ dimensional gamma matrices by taking tensor products as
\begin{equation}
\begin{array}{cc}
(\gamma^{\mu}\otimes\sigma^{1},~\gamma^{(d+1)}
\otimes\sigma^{1},~1\otimes\sigma^{2})~~~~&~~~~\mbox{:~up~to~a~factor~}i\,.
\end{array}
\label{offcon}
\end{equation}
Therefore  the gamma matrices in even dimensions can be chosen to have the
``off-block diagonal'' form
\begin{equation}
\begin{array}{cc}
\gamma^{\mu}=\left(\begin{array}{cc}
                    0&\sigma^{\mu}\\
                    \tilde{\sigma}^{\mu}&0
                    \end{array}\right)\,,~~~~&~~~~
\gamma^{(d+1)}=\left(\begin{array}{cc}
                      1&0\\
                      0&-1
                      \end{array}\right)\,,
\end{array}
\label{offdiagonal}
\end{equation}
%%%This can be shown from eq.(\ref{dd+2}) and by fixing
%%%\begin{equation}
%%%\gamma^{(d+1)}=\left(\begin{array}{cc}
%%%                      1&0\\
%%%                      0&-1
%%%                      \end{array}\right)
%%%\label{offd+1}
%%%\end{equation}
where the $2^{\frac{d}{2}-1}\times 2^{\frac{d}{2}-1}$ matrices,
$\sigma^{\mu},\tilde{\sigma}^{\mu}$ satisfy
\begin{eqnarray}
&\sigma^{\mu}\tilde{\sigma}^{\nu}+\sigma^{\nu}\tilde{\sigma}^{\mu}=2\eta^{\mu\nu}\,,&\\
&{}&\nonumber\\
&\sigma^{\mu}{}^{\dagger}=\tilde{\sigma}_{\mu}\,.&
\end{eqnarray}
In this choice of gamma matrices, from eq.(\ref{daggerA}),
$A_{\pm},B_{\pm},C_{\pm}$ are either ``block diagonal'' or ``off-block diagonal''
depending on whether $t,\,\frac{t-s}{2},\,\frac{t+s}{2}$
are even or odd respectively.\newline
In particular, in the case of odd $t$, we
write from eqs.(\ref{defA},\,\ref{AAA}) $A$ as
\begin{equation}
\begin{array}{cc}
A=\left(\begin{array}{cc}
          0& \a  \\
          \ta& 0
          \end{array}\right)\,,~~~~&~~~~
\a=\sqrt{(-1)^{\frac{t(t-1)}{2}}}\sigma^{1}\tilde{\sigma}^{2}\cdots\sigma^{t}
=\ta^{\dagger}=\ta^{-1}\,,
\end{array}
\label{offA}
\end{equation}
and in the case of odd $\frac{t+s}{2}$ we write from
eq.(\ref{Ctranspose}) $C_{\pm}$ as
\begin{equation}
\begin{array}{cc}
C_{\pm}=\left(\begin{array}{cc}
          0&\c\\
          \pm\tc& 0
          \end{array}\right)\,,~~~~&~~~~
\c=\varepsilon_{+}(-1)^{\frac{t(t-1)}{2}}\tc{}^{T}=(\c^{\dagger})^{-1}\,,
\end{array}
\label{offC}
\end{equation}
where $\a,\,\ta,\,\c,\,\tc$ satisfy from eqs.(\ref{AgA},\,\ref{CgC})
\begin{equation}
\begin{array}{ll}
\sigma^{\mu}{}^{\dagger}=\ta\sigma^{\mu}\ta\,,~~~~&~~~~
\tilde{\sigma}^{\mu}{}^{\dagger}=\a\tilde{\sigma}^{\mu}\a\,,\\
{}&{}\\
\sigma^{\mu}{}^{T}=(-1)^{t+1}\tc\sigma^{\mu}\c^{-1}\,,~~~~&~~~~
\tilde{\sigma}^{\mu}{}^{T}=(-1)^{t+1}\c\tilde{\sigma}^{\mu}\tc^{-1}\,.
\end{array}
\label{ac}
\end{equation}
If both of $t$ and $\frac{t+s}{2}$ are odd then from eq.(\ref{ABCpm})
\begin{equation}
\begin{array}{cc}
\a{}^{T}=(-1)^{\frac{t-1}{2}}\tc\,\a\,\c^{-1}\,,~~~~&~~~~
\ta{}^{T}=(-1)^{\frac{t-1}{2}}\c\,\ta\,\tc{}^{-1}\,.
\end{array}
\end{equation}

%%%%%%%%%%%%%%%%%%%%%%%%%%%%%%%%%%%%%%%%%%%%%%%%%%%%%%%%%%%%%%%%%%%%%%%%%%%%%
%%%%%%%%%%%%%%%%%%%%%%%%%%%%%%%%%%%%%%%%%%%%%%%%%%%%%%%%%%%%%%%%%%%%%%%%%%%%%
%%%%%%%%%%%%%%%%%%%%%%%%%%%%%%%%%%%%%%%%%%%%%%%%%%%%%%%%%%%%%%%%%%%%%%%%%%%%
\subsection{In Odd Dimensions}
%%%Any  representation of the Clifford algebra in odd $d+1=t+s$
%%%dimensions,  should not be smaller than
%%%$2^{d/2}\times 2^{d/2}$, since it contains
%%% subalgebras, which are  even $d$ dimensional  gamma matrices. In fact, as
%%% it is well known,
The gamma matrices in  odd $d+1=t+s$ dimensions are
constructed by combining  a set of even $d$ dimensional  gamma matrices with either
$\pm\gamma^{(d+1)}$ or
$\pm i\gamma^{(d+1)}$ depending on the
signature of even $d$ dimensions.   This way of construction is general, since
$\gamma^{(d+1)}$ serves the role of $\gamma^{d+1}$
\begin{equation}
\begin{array}{cc}
-\gamma^{\mu}=\gamma^{d+1}\gamma^{\mu}(\gamma^{d+1})^{-1}\,,~~~~&~~~~
\mbox{for~}\mu=1,2,\cdots,d\,,\\
{}\\
\multicolumn{2}{c}{(\gamma^{d+1})^{2}=\pm 1\,,}
\end{array}
\end{equation}
and such a matrix is unique in irreducible representations up to sign.\newline

However, contrary to the even dimensional Clifford algebra, in odd
dimensions two different choices of the signs in $\gamma^{d+1}$ bring
two irreducible representations for the Clifford algebra,
which can not be mapped to each other\footnote{Nevertheless,
this can be cured by the following transformation.  Under $x^{\mu}=(x^{1},x^{2},\cdots,x^{d+1})~\rightarrow~x^{\prime}{}^{\mu}
=(x^{1},x^{2},\cdots,-x^{d+1})$, we transform the Dirac field
$\psi(x)$ as $
\psi(x)~~\rightarrow~~\psi^{\prime}(x^{\prime})=\psi(x)\,,$
to get
$\bar{\psi}(x)\gamma\cdot\partial\psi(x)~~\rightarrow~~
\bar{\psi}^{\prime}(x^{\prime})\gamma^{\prime}\cdot\partial^{\prime}
\psi^{\prime}(x^{\prime})=
\bar{\psi}(x)\gamma\cdot\partial\psi(x)\,.$
Hence those two representations are  equivalent describing
the same physical system.}
 by similarity
transformations
\begin{equation}
\begin{array}{ccc}
\gamma^{\mu}=(\gamma^{1},\gamma^{2},\cdots,\gamma^{d+1})~~~~\mbox{and}~~~~
\gamma^{\prime}{}^{\mu}=(\gamma^{1},\gamma^{2},
\cdots,\gamma^{d},-\gamma^{d+1})\,.
\end{array}
\end{equation}
If there were a similarity transformation between these two, it should
have been identity up to constant because of the uniqueness of the
similarity transformation in even dimensions. Clearly this would be a
contradiction due to the presence of the two opposite signs
in $\gamma^{d+1}$. \newline
%%%%%%%%%%%%%%%%%
%%%Thus, in odd $d+1$ dimensions, $2^{d/2}\times
%%%2^{d/2}$  representations
%%%of gamma matrices are unique up to similarity and
%%%reflection.\newline
%%%%%%%%%%%%%%%

In general one can put\footnote{Our results~(\ref{indep1}-\ref{indep2})
 do not depend on the choice of  the signature in $d$
dimensions,
i.e. they hold for either increasing the time dimensions,~$d=(t-1)+s$
or the space dimensions,~$d=t+(s-1)$.}
\begin{equation}
\,\gamma^{d+1}=\left\{\begin{array}{ll}
    \pm\gamma^{12\cdots d}~&~\mbox{for~}t-s \equiv 1~\mbox{mod}~4\,, \\
{}&{}\\
           \pm i\gamma^{12\cdots d}~&~\mbox{for~}t-s\equiv 3~\mbox{mod}~4\,.
                        \end{array}\right.
\label{indep1}
\end{equation}

$2^{d/2}\times 2^{d/2}$
gamma matrices in odd $d+1$ dimensions, $\gamma^{\mu}, \mu=1,2,\cdots,d+1$,
induce the following  basis of
$2^{d/2}\times 2^{d/2}$ matrices, $\tilde{\Gamma}^{M}$
\begin{equation}
\begin{array}{cc}
\tilde{\Gamma}^{M}=(1,\gamma^{\mu},\gamma^{\mu\nu},
\cdots,\gamma^{\mu_{1}\mu_{2}\cdots\mu_{d/2}})\,,~~~&~~~M=1,2,\cdots 2^{d}\,.
\end{array}
\end{equation}
From eq.(\ref{indep1})
\begin{equation}
\begin{array}{c}
\tilde{\Gamma}^{M}\tilde{\Gamma}^{N}
=\tilde{\Omega}_{MN}\tilde{\Gamma}^{L}\,,\\
{}\\
\tilde{\Omega}_{MN}=\left\{\begin{array}{ll}
\pm 1~&\mbox{for~}t-s\equiv 1~\mbox{mod}~4\,, \\
                  {}&{}\\
              \pm 1,\pm i ~&\mbox{For~}t-s\equiv 3~\mbox{mod}~4\,.
                          \end{array}\right.
\end{array}
\end{equation}
Here, contrary to the even dimensional case,
$\tilde{\Omega}_{MN}$ depends on each
particular choice of the
representations due to the arbitrary sign factor in   $\gamma^{d+1}$.
This is why  eq.(\ref{similarity})
does  not hold in odd dimensions. Therefore it is not peculiar that
not  all of
$\pm\gamma^{\mu}{}^{\dagger},\pm\gamma^{\mu}{}^{\ast},\pm\gamma^{\mu}{}^{T}$
are  related to $\gamma^{\mu}$ by similarity transformations. In
fact,  if
it were true, say for $\pm\gamma^{\mu}{}^{\ast}$,
then  the similarity
transformation should have been
$B_{\pm}$~(\ref{B}) by the uniqueness of the
similarity transformations in even dimensions, but this would be  a
contradiction
 to eq.(\ref{daggerA}), where the sign does not alternate under the change
of $B_{+}\leftrightarrow B_{-}$. Thus, in odd dimensions,
only the half of
$\pm\gamma^{\mu}{}^{\dagger},\pm\gamma^{\mu}{}^{\ast},\pm\gamma^{\mu}{}^{T}$
are related to $\gamma^{\mu}$ by similarity transformations and hence from eq.(\ref{daggerA})
there exist three similarity transformations, $A,B,C$  such that
\begin{eqnarray}
(-1)^{t+1}\gamma^{\mu}{}^{\dagger}=A\gamma^{\mu}A^{-1}\,,\label{oddA}\\
{}\nonumber\\
(-1)^{\frac{t-s-1}{2}}\gamma^{\mu}{}^{\ast}=B\gamma^{\mu}B^{-1}\,,\label{oddB}\\
{}\nonumber\\
(-1)^{\frac{t+s-1}{2}}\gamma^{\mu}{}^{T}=C\gamma^{\mu}C^{-1}\,.\label{oddC}
\end{eqnarray}
$A,\,B,\,C$ are all unitary and satisfy
\begin{eqnarray}
A=A^{-1}=A^{\dagger}\,,~~~~~~~~~~~
C=B^{T}A\,,~~~~~~\\
{}\nonumber\\
B^{\ast}B=\varepsilon\,1=(-1)^{\frac{1}{8}(t-s+1)(t-s-1)}\,1\,,
~~~~~~~~~~\label{B*B2}\\
{}\nonumber\\
B^{T}=\varepsilon B\,,~~~~~~~~~~~~
C^{T}=\varepsilon (-1)^{\frac{ts}{2}}\,C=(-1)^{\frac{1}{8}(t+s+1)(t+s-1)}C\,,~~~~~
\label{BCT}\\
{}\nonumber\\
(-1)^{\frac{ts}{2}}A^{T}=BAB^{-1}=CAC^{-1}\,.~~~~~~~~\label{indep2}
\end{eqnarray}
In particular, $A$ is given by eq.(\ref{defA}).

%%%%%%%%%%%%%%%%%%%%%%%%%%%%%%%%%%%%%%%%%%%%%%%%%%%%%%%%%%%%%%%%%%%%%%%%%%%%%%%%%%%%%%%%%%%%%%%%%%%%%%%%%%%%%%%%%%%%%%%%%%%%%%%%%%%%%%%%%%%%%%%%%%%%%%%%%%%%%%%%%%%%%%%%%%%%%%%%%%%%%%%%%%%%%%%%%%%%%%%%%%%%%%%%%%%%%%%%%%%%%%%%%%%%%%%%%%%%%%%%%%%%%%%%%%%%%%%%%%%%%%%%%%%%%%%%%%%%%%%%%%%%%%%%%%%%%%%%%%%%%%%%%%%%%%%%%%%%%%%%%%%%%%%%%%%%%%%%%%%%%%%%%%%%%%%%%%%%%%%%%%%%%%%%%%%%%%%%%%%%%%%%%%%%%%%%%%
\subsection{Lorentz Transformations}
Lorentz transformations,~$L$ can be represented by
the following  action on gamma matrices in a standard way
\begin{equation}
\L^{-1}\gamma^{\mu}\L=L^{\mu}_{~\nu}\gamma^{\nu}\,,
\label{LgL}
\end{equation}
where $L$ and $\L$ are given by
\begin{equation}
\begin{array}{cc}
L=e^{w_{\mu\nu}M^{\mu\nu}}\,,~~~~&~~~~
\L=e^{\frac{1}{2}w_{\mu\nu}\gamma^{\mu\nu}}\,,\\
{}&{}\\
\multicolumn{2}{c}{
(M^{\mu\nu})^{\lambda}_{~\rho}=\eta^{\mu\lambda}\delta^{\nu}_{~\rho}-
\eta^{\nu\lambda}\delta^{\mu}_{~\rho}\,.}
\end{array}
\label{calL}
\end{equation}
For  even $d$, if a
$2^{d/2}\times 2^{d/2}$
matrix,~$M^{\mu_{1}\mu_{2}\cdots\mu_{n}}$,
is totally anti-symmetric over the  $n$ spacetime
indices
\begin{equation}
M^{\mu_{1}\mu_{2}\cdots\mu_{n}}=M^{[\mu_{1}\mu_{2}\cdots\mu_{n}]}\,,
\label{antisym-mu}
\end{equation}
and transforms covariantly under Lorentz transformations in $d$ or
$d+1$ dimensions as
\begin{equation}
\L^{-1}M^{\mu_{1}\mu_{2}\cdots\mu_{n}}\L=
\prod_{i=1}^{n}L^{\mu_{i}}{}_{\nu_{i}}\,M^{\nu_{1}\nu_{2}\cdots\nu_{n}}\,,
\label{covM}
\end{equation}
then for $0\leq n\leq \max (d/2,2)$, the general forms of
$M^{\mu_{1}\mu_{2}\cdots\mu_{n}}$ are
%%%\footnote{If we loose the anti-symmetric
%%%condition~(\ref{antisym-mu}), then the general forms are the products of
%%%the metric,~$\eta^{\mu\nu}$, and the forms in eq.(\ref{Mforms})
%%%in an obvious
%%%way, $\sum_{n}\eta^{n}\gamma^{\mu_{2n+1}\mu_{2n+2}\cdots}$.}
\begin{equation}
M^{\mu_{1}\mu_{2}\cdots\mu_{n}}=\left\{\begin{array}{ll}
(1+c\gamma^{(d+1)})\gamma^{\mu_{1}\mu_{2}\cdots\mu_{n}}~~&~~~
\mbox{In~even~$d$~dimensions\,,}\\
{}&{}\\
\gamma^{\mu_{1}\mu_{2}\cdots\mu_{n}}~~&~~~\mbox{In~odd~$d+1$~dimensions\,,}
\end{array}\right.
\label{Mforms}
\end{equation}
where $c$ is a constant.\newline

To show this, one may first expand  $M^{\mu_{1}\mu_{2}\cdots\mu_{n}}$ in
terms of $\gamma_{\nu_{1}\nu_{2}\cdots\nu_{m}},\,
\gamma^{(d+1)}\gamma_{\nu_{1}\nu_{2}\cdots\nu_{m}}$ or
$\gamma_{\nu_{1}\nu_{2}\cdots\nu_{m}}$ depending on the
dimensions,~$d$ or $d+1$, with $0\leq m\leq d/2$.
Then eq.(\ref{covM}) implies that
the coefficients
of them, say $T^{\mu_{1}\mu_{2}\cdots\mu_{m+n}}$,
 are Lorentz invariant tensors satisfying
\begin{equation}
\prod^{m+n}_{i=1}L^{\mu_{i}}_{~\nu_{i}}\,T^{\nu_{1}\nu_{2}\cdots\nu_{m+n}}=
T^{\mu_{1}\mu_{2}\cdots\mu_{m+n}}
\label{LI}
\end{equation}
Finally one can recall  the well known fact~\cite{weyl} that
the general forms of Lorentz invariant
tensors are multi-products of the metric,\,$\eta^{\mu\nu}$, and the totally
antisymmetric
tensor,\,$\epsilon^{\mu_{1}\mu_{2}\cdots}$, which verifies eq.(\ref{Mforms}).

%%%%%%%%%%%%%%%%%%%%%%%%%%%%%%%%%%%%%%%%%%%%%%%%%%%%%%%%%%%%%%%%%%%%%%%%%%%%%%%%%%%%%%%%%%%%%%%%%%%%%%%%%%%%%%%%%%%%%%%%%%%%%%%%%%%%%%%%%%%%%%%%%%%%%%%%%%%%%%%%%%%%%%%%%%%%%%%%%%%%%%%%%%%%%%%%%%%%%%%%%%%%%%%%%%%%%%%%%%%%%%%%%%%%%%%%%%%%%
\subsection{Crucial Identities for Super Yang-Mills}
The following identities are crucial to show the existence of the non-Abelian  super Yang-Mills in THREE, FOUR, SIX and TEN dimensions.\newline
\newline
\newline
(i) The following identity holds only in THREE or FOUR dimensions with arbitrary signature
\begin{equation}
0=(\gamma^{\mu}C^{-1})_{\alpha\beta}(\gamma_{\mu}C^{-1})_{\gamma\delta}+
\mbox{cyclic~permutations~of~}\alpha,\beta,\gamma
\label{gammaper}
\end{equation}
To verify the identity in even dimensions we contract
$(\gamma^{\mu}C^{-1})_{\alpha\beta}(\gamma_{\mu})_{\gamma\delta}$
with $(C\gamma^{\nu_{1}\nu_{2}\cdots\nu_{n}})_{\beta\alpha}$ and take
cyclic permutations of $\alpha,\beta,\gamma$ to get
\begin{equation}
0=2^{d/2}\delta^{n}_{1}+(d-2n)(\zeta+
\zeta^{n}(-1)^{\frac{1}{2}n(n-1)})(-1)^{n+\frac{1}{8}d(d-\zeta 2)}
\end{equation}
This equation must be   satisfied  for all $0\leq n\leq d$,
which is valid  only in $d=4,\zeta=-1$. \newline
Similar analysis can be done
 for the $d+1$ odd dimensions  by adding
$(\gamma^{(d+1)}C^{-1})_{\alpha\beta}(\gamma^{(d+1)}C^{-1})_{\gamma\delta}$
term into eq.(\ref{gammaper}).
We get
\begin{equation}
\begin{array}{cc}
0=2^{d/2}(\delta^{n}_{1}+\delta^{n}_{d})+(d-2n+1)(\zeta+
\zeta^{n}(-1)^{\frac{1}{2}n(n-1)})(-1)^{n+\frac{1}{8}d(d-\zeta
2)},~~~&~~~
\zeta=(-1)^{d/2}
\end{array}
\end{equation}
Only in $d=2$ and hence three dimensions, this equation is satisfied for all $0\leq n\leq d$.\newline
\newline
\newline
(ii)  The following identity holds only in TWO, FOUR or SIX dimensions with arbitrary signature
\begin{equation}
0=(\sigma^{\mu})_{\alpha\beta}(\sigma_{\mu})_{\gamma\delta}+
(\sigma^{\mu})_{\gamma\beta}(\sigma_{\mu})_{\alpha\delta}
\label{symag}
\end{equation}
To verify this identity  we take  $d$
dimensional sigma matrices  from $f=d-2$ dimensional gamma matrices as
in  eq.(\ref{offcon})
\begin{equation}
\sigma^{\mu}=(\gamma^{\mu},\gamma^{(f+1)},i)
\end{equation}
to get
\begin{equation}
(\sigma^{\mu})_{\alpha\beta}(\sigma_{\mu})_{\gamma\delta}=
(\gamma^{\mu})_{\alpha\beta}(\gamma_{\mu})_{\gamma\delta}+
(\gamma^{(f+1)})_{\alpha\beta}(\gamma^{(f+1)})_{\gamma\delta}
-\delta_{\alpha\beta}\delta_{\gamma\delta}
\end{equation}
Again this expression is valid for any signature, $(t,s)$. Now we
contract this equation with $(\gamma^{\nu_{1}\nu_{2}\cdots\nu_{n}}
C_{+}^{-1})_{\beta\delta}$. From
eqs.(\ref{CgC},\,\ref{daggerA})
in the case of odd $t$ we get
\begin{equation}
\left((-1)^{n}(f-2n)+(-1)^{\frac{f}{2}+n}-1\right)
(\gamma^{\nu_{1}\nu_{2}\cdots\nu_{n}}C_{+}^{-1})_{\alpha\gamma}
\end{equation}
To satisfy eq.(\ref{symag}) this expression must be anti-symmetric
over $\alpha\leftrightarrow\gamma$ for any $0\leq n\leq f$.
Thus from eq.(\ref{transpose}) we must require
$0=(-1)^{n}(f-2n)+(-1)^{\frac{f}{2}+n}-1$ for all $n$ satisfying
 $(-1)^{\frac{1}{8}f(f-2)+\frac{1}{2}n(n-1)}=1$. This condition is
satisfied only in $f=0,2,4$ and hence $d=2,4,6$~~($f=6$ case is excluded by choosing $n=6$
and $f\geq 8$ cases are excluded by choosing either $n=0$ or
$n=3$).\newline
\newline
\newline
(iii) The following identity holds only in TWO or TEN dimensions with arbitrary signature
\begin{equation}
0=(\sigma^{\mu}\c^{-1})_{\alpha\beta}(\sigma_{\mu}\c^{-1})_{\gamma\delta}+
\mbox{cyclic~permutations~of~}\alpha,\beta,\gamma
\end{equation}

\newpage

%%%%%%%%%%%%%%%%%%%%%%%%%%%%%%%%%%%%%%%%%%%%%%%%%%%%%%%%%%%%%%%%%%%%%%%%%%%%
%%%%%%%%%%%%%%%%%%%%%%%%%%%%%%%%%%%%%%%%%%%%%%%%%%%%%%%%%%%%%%%%%%%%%%%%%%%%
%%%%%%%%%%%%%%%%%%%%%%%%%%%%%%%%%%%%%%%%%%%%%%%%%%%%%%%%%%%%%%%%%%%%%%%%%%%
\section{Spinors}
\subsection{Weyl Spinor}
In any even $d$ dimensions, Weyl spinor,~$\psi$, satisfies
\begin{equation}
\gamma^{(d+1)}\psi=\psi
\label{weyl1}
\end{equation}
and so ${\bar{\psi}}={\psi^{\dagger}}A$ satisfies from eq.(\ref{daggerA})
\begin{equation}
\begin{array}{cc}
\bar{\psi}\gamma^{(d+1)}=(-1)^{t}\bar{\psi}~~~~&~~~~
\gamma^{(d+1)}C_{\pm}^{-1}\bar{\psi}^{T}=(-1)^{\frac{t-s}{2}}C_{\pm}^{-1}
\bar{\psi}^{T}
\end{array}
\label{weyl2}
\end{equation}

%%%%%%%%%%%%%%%%%%%%%%%%%%%%%%%%%%%%%%%%%%%%%%%%%%%%%%%%%%%%%%%%%%%%%%%%%%%%
%%%%%%%%%%%%%%%%%%%%%%%%%%%%%%%%%%%%%%%%%%%%%%%%%%%%%%%%%%%%%%%%%%%%%%%%%%%%
%%%%%%%%%%%%%%%%%%%%%%%%%%%%%%%%%%%%%%%%%%%%%%%%%%%%%%%%%%%%%%%%%%%%%%%%%%%
\subsection{Majorana Spinor}
By definition Majorana spinor  satisfies
\begin{equation}
\begin{array}{ccc}
\bar{\psi}=\psi^{T}C_{\pm}~~~~&~\mbox{or}~&~~~~\bar{\psi}=\psi^{T}C
\label{Mcon}
\end{array}
\end{equation}
depending on the dimensions, even or odd.
This is possible only if $\varepsilon_{\pm},\varepsilon=1$ and so  from
eqs.(\ref{B*B},\,\ref{B*B2})
\begin{equation}
\begin{array}{ll}
\eta=+1~:&~~~t-s=0,1,2\mbox{~mod~}8\\
{}&{}\\
\eta=-1~:~&~~~t-s=0,6,7\mbox{~mod~}8
\end{array}
\label{etats}
\end{equation}
where $\eta$ is the sign factor,~$\pm 1$, occuring in eq.(\ref{B}) or
eq.(\ref{oddB})\footnote{In~\cite{kugotownsend}, $\eta=-1$ case is called
Majorana and $\eta=+1$ case is called pseudo-Majorana.}. \newline

%%%%%%%%%%%%%%%%%%%%%%%%%%%%%%%%%%%%%%%%%%%%%%%%%%%%%%%%%%%%%%%%%%%%%%%%%
%%%%%%%%%%%%%%%%%%%%%%%%%%%%%%%%%%%%%%%%%%%%%%%%%%%%%%%%%%%%%%%%%%%%%%%
%%%%%%%%%%%%%%%%%%%%%%%%%%%%%%%%%%%%%%%%%%%%%%%%%%%%%%%%%%%%%%%%%%%%%%%%%
%%%%%%%%%%%%%%%%%%%%%%%%%%%%%%%%%%%%%%%%%%%%%%%%%%%%%%%%%%%%%%%%%%%%%%%
\subsection{Majorana-Weyl Spinor}
Majorana-Weyl spinor satisfies both of the two conditions above
\begin{equation}
\begin{array}{cc}
\gamma^{(d+1)}\psi=\psi~~~~&~~~~\bar{\psi}=\psi^{T}C_{\pm}
\end{array}
\end{equation}
Majorana-Weyl Spinor exists only if
\begin{equation}
\begin{array}{ll}
\eta=+1~:&~~~t-s=0\mbox{~mod~}8\\
{}&{}\\
\eta=-1~:~&~~~t-s=0\mbox{~mod~}8
\end{array}
\end{equation}
\newpage

%%%%%%%%%%%%%%%%%%%%%%%%%%%%%%%%%%%%%%%%%%%%%%%%%%%%%%%%%%%%%%%%%%%%%%%%%%%%%
%%%%%%%%%%%%%%%%%%%%%%%%%%%%%%%%%%%%%%%%%%%%%%%%%%%%%%%%%%%%%%%%%%%%%%%%%%%%%%%%%%%%
\section{Majorana Representation and $\mbox{SO}(8)$}
{\bf Fact 1:}\\ Consider a finite dimensional vector space, ${\cal V}$ with the unitary and symmetric  matrix, $\B=\B^{T}$, $\B\B^{\dagger}=1$. For
every $|v\rangle\in{\cal V}$ if $\B|v\rangle^{\ast}\in{\cal V}$ then there exists an orthonormal ``semi-real '' basis, ${\cal
V}=\{|l\rangle,\,l=1,2,\cdots\}$ such that
$\B|l\rangle^{\ast}=|l\rangle$.\\
\\
\textit{Proof}\\ Start with an arbitrary orthonormal bais, $\{|v_{l}\rangle,\,l=1,2,\cdots\}$ and let
$|1\rangle\propto|v_{1}\rangle+\B|v_{1}\rangle^{\ast}$.   After  the normalization, $\langle 1|1\rangle=1$, we can take a new orthonormal basis,
$\{|1\rangle, |2^{\prime}\rangle, |3^{\prime}\rangle,\cdots\}$.  Now we assume that $\{|1\rangle, |2\rangle, \cdots|k-1\rangle,|k^{\prime}\rangle,
|(k+1)^{\prime}\rangle, \cdots\}$ is an orhonormal basis such that  $\B|j\rangle^{\ast}=|j\rangle$  for $1\leq j\leq k-1$. To construct the $k$\,th
such a vector, $|k\rangle$ we set $|k\rangle\propto|k^{\prime}\rangle+\B|k^{\prime}\rangle^{\ast}$ with  the normalization. We check this is
orthogonal to $|j\rangle,\,1\leq j\leq k-1$
\begin{equation}
\langle j|\displaystyle{\left(\,|k^{\prime}\rangle+\B|k^{\prime}\rangle^{\ast}\,\right)}=0+\langle
 k|j\rangle=0\,.
\end{equation}
In this way one can construct the desired basis.\\

In the spacetime which admits Majorana spinor  from Eq.(\ref{etats})
\begin{equation}
\begin{array}{ll}
\eta=+1~:&~~~t-s=0,1,2\mbox{~mod~}8\\
{}&{}\\
\eta=-1~:~&~~~t-s=0,6,7\mbox{~mod~}8\,,
\end{array}
\end{equation}
more explicitly in the even dimensions having $\varepsilon_{+}=1$ (or $\varepsilon_{-}=1$) where  $B_{+}$ (or $B_{-}$) is symmetric and also in the
odd dimensions of $\varepsilon=1$ where $B$ is symmetric, from  {\bf Fact 1} above we can choose an ``semi-real '' orthonormal basis such that
$B_{\eta}^{-1}|l\rangle^{\ast}=|l\rangle$ (here it is  $B_{\eta}^{-1}$ that plays the role of $\B$ in {\bf Fact 1}). In the basis, we write the gamma matrices
\begin{equation}
\gamma^{\mu}=\sum\,R^{\mu}_{lm}|l\rangle\langle m|\,.
\end{equation}
From $\eta\,\gamma^{\mu}{}^{\ast}=B_{\eta}\gamma^{\mu}B_{\eta}^{-1}$ and the property of the semi-real basis, $B_{\eta}^{-1}|l\rangle^{\ast}=|l\rangle$ we
get
\begin{equation}
\big( R^{\mu}_{lm}\big)^{\ast}=\eta R^{\mu}_{lm}\,.\label{Reta1}
\end{equation}
Since $R^{\mu}$ is also a representation of the gamma matrix
\begin{equation}
R^{\mu}R^{\nu}+R^{\nu}R^{\mu}=2\eta^{\mu\nu}\,,\label{Reta2}
\end{equation}
adopting the true real basis, we conclude that \textbf{there exists a Majorana represention where the gamma matrices are real, $\eta=+$ or pure imaginary,
$\eta=-$ in any spacetime admitting Majorana spinors.}\\

Furthermore from Eq.(\ref{daggerA}), in the even dimension of $t-s\equiv 0$ mod $8$, $\varepsilon_{\pm}=1$ and
$\gamma^{(d+1)}{}^{\ast}=B\gamma^{(d+1)}B^{-1}$ (here we omit the subscript index $\pm$ or $\eta$      for simplicity.). The action,
$|v\rangle\rightarrow B^{\dagger}|v\rangle^{\ast}$ preserves the chirality, and from  the {\bf fact 1} above we can choose an orthonormal semi-real
basis for the chiral and anti-chiral spinor spaces, ${\cal V}={\cal V}_{+}+{\cal V}_{-}$, ${\cal V}_{\pm}=\{|l_{\pm}\rangle\}$ such that
\begin{equation}
\begin{array}{cccc}
\langle l_{\pm}|m_{\pm}\rangle=\delta_{lm}\,,~~&~~\langle l_{\pm}|m_{\mp}\rangle=0\,,~~&~~\gamma^{(d+1)}|l_{\pm}\rangle=\pm|l_{\pm}\rangle\,,
~~&~~B^{\dagger}|l_{\pm}\rangle^{\ast}=|l_{\pm}\rangle\,.
\end{array}
\label{reality}
\end{equation}
With the semi-real basis
\begin{equation}
\gamma^{(d+1)}=\left(\begin{array}{cc}1&0\\0&-1\,\end{array}\right)\,,
\end{equation}
and the gamma matrices are in the Majorana representation
\begin{equation}
\begin{array}{ccc}
\gamma^{\mu}=\left(\begin{array}{cc}0&r^{\mu}\\r_{\mu}^{T}&0\,\end{array}\right)\,,~~&~~r^{\mu}\in\mbox{O}(2^{d/2-1})\,,~~&~~
r^{\mu}r^{\nu}{}^{T}+r^{\nu}r^{\mu}{}^{T}=2\delta^{\mu\nu}\,.
\end{array}
\end{equation}

From Eq.(\ref{reality}) any two sets of semi-real basis, say $\{|l_{\pm}\rangle\}$ and $\{|\tilde{l}_{\pm}\rangle\}$ are connected by an
$\mbox{O}((2^{d/2-1}))$ transformation
\begin{equation}
\begin{array}{cc}
|\tilde{l}_{\pm}\rangle=\displaystyle{\sum_{m}}\,\Lambda_{\pm ml}|m_{\pm}\rangle\,,~~&~~\displaystyle{\sum_{m}}\,\Lambda_{\pm lm}\Lambda_{\pm
nm}=\delta_{ln}\,.
\end{array}
\end{equation}
If we define
\begin{equation}
\Lambda_{\pm}=\displaystyle{\sum_{l,m}}\,\Lambda_{\pm lm}|l_{\pm}\rangle\langle m_{\pm}|\,,
\end{equation}
then $|\tilde{l}_{\pm}\rangle=\Lambda_{\pm}|l_{\pm}\rangle$ and from the definition of the semi-real basis
\begin{equation}
\begin{array}{cc}
\Lambda_{\pm}=B^{\dagger}\Lambda_{\pm}^{\ast}B=\Lambda_{\pm}P_{\pm}=P_{\pm}\Lambda_{\pm}\,, ~~&~~\Lambda_{\pm}\Lambda_{\pm}^{\dagger}=P_{\pm}\,.
\end{array}
\end{equation}
We write
\begin{equation}
\begin{array}{cc}
\displaystyle{\Lambda_{\pm}=e^{M_{\pm}}}\,,~~&~~
M_{\pm}\equiv\displaystyle{\sum_{n=1}^{\infty}}\,(-1)^{n+1}\frac{1}{n}(\Lambda_{\pm}-P_{\pm})^{n}=\ln\,\Lambda_{\pm}\,.
\end{array}
\end{equation}
Thus for $\Lambda_{\pm}$ such that the infinity sum converges we have
\begin{equation}
M_{\pm}=-M_{\pm}^{\,\dagger}=B^{\dagger}M_{\pm}^{\,\ast}B=M_{\pm}P_{\pm}=P_{\pm}M_{\pm}\,.
\end{equation}
This gives a strong constraint when we  express $M_{\pm}$ by the gamma matrix products. For the Eucledean eight dimensions only the $\mbox{SO}(8)$
generators for the spinors survive in the expansion!
\begin{equation}
M_{\pm}=\textstyle{\frac{1}{2}}w_{ab}\gamma^{ab}P_{\pm}\,.
\end{equation}
Namely we find an isomorphism between the two $\mbox{SO}(8)$'s, one for the semi-real vectors and the other for the spinors in the conventional
sense. Alternatively this can be seen from
\begin{equation}
\gamma^{ab}=\left(\begin{array}{cc}r^{[a}r^{b]\,T}&0\\0&r^{[a\,T}r^{b]}\end{array}\right)\,,
\end{equation}
where the each block diagonal is a generator of $\mbox{SO}(D)$ while the dimension of the chiral space is
$2^{d/2-1}$. Only in $d=8$ both coincide leading to the ``$\mbox{so}(8)$ triolity" among $\mbox{so}_{v}(8)$, $\mbox{so}_{c}(8)$ and $\mbox{so}_{\bar{c}}(8)$.\\ ~\\

{\bf Fact 2:} Relation to octonions.\\
In Euclidean eight dimensions, the $16\times 16$ gamma matrices can be taken of the off-block diagonal form,
\begin{equation}
\begin{array}{cc}
\gamma_{a}=\left(\begin{array}{cc}
                    0&r_{a}\\
                    r^{T}_{a}&0
                    \end{array}\right)\,,~~~~&~~~~
r_{a}r_{b}^{T}+r_{b}r_{a}^{T}=2\delta_{ab}\,,
\end{array}
\end{equation}
 where the $8\times 8$ real matrices, $r_{a}$,  $1\leq a\leq 8$, give the multiplication of the  octonions, $o_{a}$,
\be
o_{a}o_{b}=(r_{a})_{b}{}^{c}o_{c}\,.
\ee
~\\

{\bf Fact 3:}\\ Consider an arbitrary real self-dual or anti-self-dual four form in $D=8$
\begin{equation}
T^{\pm}_{abcd}=\pm\textstyle{\frac{1}{4!}}\,\epsilon_{abcdefgh}T^{\pm efgh}\,.
\end{equation}
Using the $\mbox{SO}(8)$ rotations one can transform the four form into the canonical form where
%%%
%%, with the notation
%%$\{(r,r^{\prime}),(s,s^{\prime}),(t,t^{\prime}),(u,u^{\prime})\}=\{(1,2),(3,4),(5,6),(7,8)\}$,
%%%
the non-vanishing components are
$T^{\pm}_{1234},\,T^{\pm}_{1256},\,T^{\pm}_{1278},\,T^{\pm}_{1357},\,T^{\pm}_{1368},\,T^{\pm}_{1458},\,T^{\pm}_{1467}$
and their dual counter parts only.\\
\\
\textit{Proof}\\ We start with the seven  linearly independent traceless Hermitian matrices
\begin{equation}
\begin{array}{cccc}
E_{\pm 1}=\gamma^{2341}P_{\pm}\,,~&E_{\pm 2}=\gamma^{2561}P_{\pm}\,,~&E_{\pm 3}=\gamma^{2781}P_{\pm}\,,~& E_{\pm 4}=\gamma^{1357}P_{\pm}\,,\\
{}&{}&{}&{}\\ E_{\pm 5}=\gamma^{3681}P_{\pm}\,,~&E_{\pm 6}=\gamma^{4581}P_{\pm}\,,~& E_{\pm 7}=\gamma^{4671}P_{\pm}\,.&{}
\end{array}
\end{equation}
As they commute with each other,  there exists a basis $ {\cal V}_{\pm}=\{|l_{\pm}\rangle\}$ diagonalizing the seven quantities
\begin{equation}
\begin{array}{ll}
E_{\pm r}=\displaystyle{\sum_{l}}\,\lambda_{rl}|l_{\pm}\rangle\langle l_{\pm}|\,,~~&~~(\lambda_{rl})^{2}=1\,.
\end{array}
\end{equation}
Further, since $C|l_{\pm}\rangle^{\ast}$ is also an eigenvector of the same eigenvalues,  from the {\bf fact 1} we
can impose the semi-reality condition without loss of generality, $C|l_{\pm}\rangle^{\ast}=|l_{\pm}\rangle$.\\

Now for the self-dual four form we let
\begin{equation}
T^{\pm}=\textstyle{\frac{1}{4\!}}\,T^{\pm}_{abcd}\gamma^{abcd}\,.
\end{equation}
Since $T^{\pm}$ is Hermitian and $C(T^{\pm})^{\ast}C^{\dagger}=T^{\pm}$, one can diagonalize $T^{\pm}$ with a semi-real basis
\begin{equation}
\begin{array}{cc}
T^{\pm}=\displaystyle{\sum_{l}}\,\lambda_{l}|\tilde{l}_{\pm}\rangle\langle \tilde{l}_{\pm}|\,,
~~&~C|\tilde{l}_{\pm}\rangle^{\ast}=|\tilde{l}_{\pm}\rangle\,.
\end{array}
\end{equation}

For the two semi-real basis above we define a transformation matrix
\begin{equation}
O_{\pm}= |l_{\pm}\rangle\langle\tilde{l}_{\pm}|\,.
\end{equation}
Then, since $T^{\pm}$ is traceless, $O_{\pm}T^{\pm}O_{\pm}^{\,\dagger}$ can be written in terms of $E_{\pm i}$'s.
Finally the fact $O_{\pm}$ gives a spinorial $\mbox{SO}(8)$ rotation   completes our proof.\\

Some useful formulae are
\begin{equation}
\begin{array}{ll}
\pm P_{\pm}&=E_{\pm 1}E_{\pm 2}E_{\pm 3}=E_{\pm 1}E_{\pm 4}E_{\pm 5}=E_{\pm 1}E_{\pm 6}E_{\pm 7}=E_{\pm 2}E_{\pm 4}E_{\pm 6}\\ {}&{}\\ {}&=E_{\pm
2}E_{\pm 5}E_{\pm 7}=E_{\pm 3}E_{\pm 4}E_{\pm 7}=E_{\pm 3}E_{\pm 5}E_{\pm 6}\,.
%%%
%%\multicolumn{2}{c}{P_{\pm}=E_{\pm i}^{2}\,.}
%%%
\end{array}
\label{EEE}
\end{equation}

For an arbitrary self-dual or anti-self-dual four form tensor in $D=8$, from
\begin{equation}
\begin{array}{ll}
T^{\pm}_{acde}T^{\pm bcde}&=(\textstyle{\frac{1}{4!}})^{2}\epsilon_{acdefghi}\epsilon^{bcdejklm}T^{\pm fghi}T^{\pm}_{jklm}\\ {}&{}\\
{}&=\textstyle{\frac{1}{4}}\,\delta_{a}{}^{b}T^{\pm}_{cdef}T^{\pm cdef}-T^{\pm}_{acde}T^{\pm bcde}\,,
\end{array}
\end{equation}
we obtain an identity
\begin{equation}
T^{\pm}_{acde}T^{\pm bcde}=\textstyle{\frac{1}{8}}\,\delta_{a}{}^{b}\,T^{\pm}_{cdef}T^{\pm cdef}\,. \label{Fidentity}
\end{equation}
\newpage

%%%%%%%%%%%%%%%%%%%%%%%%%%%%%%%%%%%%%%%%%%%%%%%%%%%%%%%%%%%%%%%%%%%%%%%%%%%%%%%%%%%%%%%%%%%%%%%%%%%%%%%%%%%%%%%%%%%%%%%%%%%%%%%%%%%%%%%%%%%%%%%%%%%%%%%%%%%%%%%%%%%%%%%%%%%%%%%%%%%%%%%%%%%%%%%%%%%%%%%%%%%%%%%%%%%%%%%%%%%%%%%%%%%%
\section{Superalgebra\label{dgraded}}
\subsection{Graded Lie Algebra}
Supersymmetry algebra is a $\Z_{2}$
graded Lie algebra, ${\mathbf{g}}=\{T_{a}\}$, which  is
an algebra with commutation and anti-commutation relations~\cite{smf,cornwell3}
\begin{equation}
[T_{a},T_{b}\}=C_{ab}^{c}T_{c}
\end{equation}
where $C_{ab}^{c}$ is the structure constant and
\begin{equation}
[T_{a},T_{b}\}=T_{a}T_{b}-(-1)^{\# a \# b}T_{b}T_{a}
\end{equation}
with $\# a$, the  $\Z_{2}$ grading of $T_{a}$,
\begin{equation}
\#a=\left\{
\begin{array}{ll}
0 & \mbox{for bosonic $a$} \\
1 & \mbox{for fermionic $a$}
\end{array}
\right.
\end{equation}
The generalized Jacobi identity is
\begin{equation}
[T_{a},[T_{b},T_{c}\}\}-(-1)^{\# a\# b}[T_{b},[T_{a},T_{c}\}\}
=[[T_{a},T_{b}\},T_{c}\}
\end{equation}
which implies
\begin{equation}
(-1)^{\# a\# c}C_{ab}^{d}C_{dc}^{e}+(-1)^{\# b\#
a}C_{bc}^{d}C_{da}^{e}+(-1)^{\# c\# b}C_{ca}^{d}C_{db}^{e}=0
\end{equation}
For  a graded Lie algebra  we consider
\begin{equation}
g(z)=\mbox{exp}(z^{a}T_{a})
\end{equation}
where $z^{a}{}$ is a superspace coordinate component which has
the same bosonic or fermionic property as $T_{a}$ and hence
$z^{a}T_{a}$ is bosonic. \newline
In the general case of non-commuting objects, say $A$ and $B$,
the Baker-Campbell-Haussdorff formula gives
\begin{equation}
\displaystyle{e^{A}e^{B}=\mbox{exp}
\left(\sum_{n=0}^{\infty}C_{n}(A,B)\right)}
\label{dBCH}
\end{equation}
where $C_{n}(A,B)$ involves $n$ commutators. The first
three of these are
\begin{equation}
\begin{array}{l}
C_{0}(A,B)=A+B\\
{}\\
C_{1}(A,B)=\textstyle{\frac{1}{2}}[A,B]\\
{}\\
C_{2}(A,B)=\textstyle{\frac{1}{12}}[[A,B],B]
+\textstyle{\frac{1}{12}}[A,[A,B]]
\end{array}
\end{equation}
Since for the graded algebra
\begin{equation}
[z^{a}T_{a},z^{b}T_{b}]=z^{b}z^{a}[T_{a},T_{b}\}=z^{b}z^{a}C_{ab}^{c}T_{c}
\end{equation}
the Baker-Campbell-Haussdorff formula~(\ref{dBCH}) implies that
$g(z)$ forms a group, the graded Lie group.
Hence we may
define a function on superspace, $f^{a}(w,z)$,  by
\begin{equation}
g(w)g(z)=g(f(w,z))
\end{equation}
Since  $g(0)=e$, the identity, we have
$f(0,z)=z,\,f(w,0)=w$ and further we
assume that $f(w,z)$ has a Taylor expansion
in the neighbourhood of $w=z=0$.  \newline
Associativity of the group multiplication requires $f(w,z)$ to satisfy
\begin{equation}
f(f(u,w),z)=f(u,f(w,z))
\label{dproperty}
\end{equation}
%%%%%%%%%%%%%%%%%%%%%%%%%%%%%%%%%%%%%%%%%%%%%%%%%%%%%%%%%%%%%
%%%Superfield, $\Psi(z)$, on superspace with coordinates, $z^{a}$, may be
%%%defined by
%%%\begin{equation}
%%%\Psi(z)=g(z)\Psi(0)g(z)^{-1}
%%%\label{dsuperfield}
%%%\end{equation}
%%%$\Psi(z)$ may carry some spin indices which we omit here for
%%%simplicity.
%%%%%%%%%%%%%%%%%%%%%%%%%%%%%%%%%%%%%%%%%%%%%%%%%%%%%%%%%%%%%%%%%%%%%%%%%%%%%%%%%%%%%%%%%%%%%%%%%%%%%%%%%%%%%%%%%%%%%%%%%%%%%%%%%%%%%%%%%%%%%%%%%%%%%%%%
\subsection{Left \& Right Invariant Derivatives}
For a graded Lie group,
left and right invariant derivatives, $L_{a},\,R_{a}$ are
defined by
\begin{eqnarray}
L_{a}g(z)=g(z)T_{a}\label{defleft}\\
{}\nonumber \\
R_{a}g(z)=-T_{a}g(z)\label{defright}
\end{eqnarray}
Explicitly we have
\begin{eqnarray}
&L_{a}=L_{a}{}^{b}(z)\partial_{b}&~~~~~~~
L_{a}{}^{b}(z)=\displaystyle{
\left.\frac{\partial f^{b}(z,u)}{\partial
u^{a}}\right|_{u=0}}\\
{}\nonumber \\
&R_{a}=R_{a}{}^{b}(z)\partial_{b}&~~~~~~~
R_{a}{}^{b}(z)=\displaystyle{
-\left.\frac{\partial f^{b}(u,z)}{\partial
u^{a}}\right|_{u=0}}
\end{eqnarray}
where~~  $\partial_{b}=\frac{\partial~}{\partial z^{b}}$.\newline
It is easy to see that $L_{a}$ is invariant under left action,
$g(z)\rightarrow hg(z)$, and  $R_{a}$ is invariant under right
action,  $g(z)\rightarrow g(z)h$.\newline
%%%%%%%%%%%%%%
%%%\begin{equation}
%%%\begin{array}{ll}
%%%L_{a}(f(w,z))=L_{a}(z)&\mbox{~~~~~left invariant}\\
%%%{}\\
%%%R_{a}(f(z,w))=R_{a}(z)&\mbox{~~~~~right invariant}\\
%%%\end{array}
%%%\end{equation}
%%%Acting on superfields, $\Psi(z)$,
%%%defined by eq.(\ref{dsuperfield}), they satisfy
%%%\begin{eqnarray}
%%%&L_{a}(z)\Psi(z)=\Psi_{a}(z)&\\
%%%{}\nonumber\\
%%%&R_{a}(z)\Psi(z)=-[T_{a},\Psi(z)\}\label{dright}&
%%%\end{eqnarray}
%%%where
%%%\begin{equation}
%%%\Psi_{a}(z)\equiv g(z)[T_{a},\Psi(0)\}g(z)^{-1}
%%%\end{equation}
%%%Thus, the action of the left invariant derivative on a superfield
%%%generates a new superfield, in the sense of eq.(\ref{dsuperfield}),
%%%while the right invariant derivative does not in general.\newline
%%%%%%%%%%%%%%%%%%%%%%%%%%%%
%%%Using
%%%$\partial_{a}\partial_{b}=(-1)^{\# a\# b}\partial_{b}\partial_{a}$ and
%%%\begin{eqnarray}
%%%&\partial_{a}=\displaystyle{
%%%\left.\frac{\partial f^{b}(-u,z)}{\partial
%%%z^{a}}\right|_{u=z}L_{b}(z)=\left.\frac{\partial f^{b}(z,-u)}{\partial
%%%z^{a}}\right|_{u=z}R_{b}(z)}&\\
%%%{}\nonumber\\
%%%&\displaystyle{
%%%\left.
%%%\frac{\partial^{2} f^{c}(u,w)}{\partial u^{a}\partial
%%%v^{b}}\right|_{u=v=0}
%%%-(-1)^{\#a\# b}
%%%\left.\frac{\partial^{2} f^{c}(u,w)}{\partial u^{b}\partial v^{a}}
%%%\right|_{u=v=0}=C_{ab}^{c}}&
%%%\end{eqnarray}
From eqs.(\ref{defleft},\,\ref{defright}) we get
\begin{eqnarray}
&[L_{a},L_{b}\}=C^{c}_{ab}L_{c}~&\\
{}\nonumber\\
&[R_{a},R_{b}\}=C^{c}_{ab}R_{c}&
\end{eqnarray}
and from  eqs.(\ref{defleft},\,\ref{defright}) we can also easily show
\begin{equation}
{[}L_{a},R_{b}\}=0
\label{dLRcom}
\end{equation}
%%%%%%%%%%%%%%
%%%Acting on superfields, these  are  consistent with
%%%\begin{eqnarray}
%%%&L_{a}(z)L_{b}(z)\Psi(z)
%%%=\displaystyle{\left.
%%%\frac{\partial^{2}~~}{\partial u^{a}\partial v^{b}}\right|_{u=v=0}}
%%%\Psi(f(f(z,u),v))
%%%=\Psi_{ab}(z)&\\
%%%{}\nonumber\\
%%%&R_{a}(z)R_{b}(z)\Psi(z)
%%%=\displaystyle{\left.
%%%\frac{\partial^{2}~~}{\partial v^{a}\partial u^{b}}\right|_{u=v=0}}
%%%\Psi(f(u,f(v,z))
%%%=[T_{b},[T_{a},\Psi(z)\}\}&
%%%\end{eqnarray}
%%%where
%%%\begin{equation}
%%%\Psi_{ab}(z)\equiv g(z)[T_{a},[T_{b},\Psi(0)\}\}g(z)^{-1}
%%%\end{equation}
Thus, $L_{a}(z)$, $R_{a}(z)$ form representations of the graded Lie
algebra separately.  For the supersymmetry algebra, the
left invariant derivatives become covariant derivatives, while
the right invariant derivatives become the generators of the
supersymmetry algebra acting on superfields.
%%%%%%%%%%%%%%%%%%%%%%%%%%%%%%%%%%%%%%%%%%%%%%%%%%%%%%%%%%%%%%%%%%%%%%%%%%%%%%%%%%%%%%%%%%%%%%%%%%%%%%%%%%%%%%%%%%%%%%%%%%%%%%%%%%%%%%%%%%%%%%%%%%%%%%%%
\subsection{Superspace \& Supermatrices}
In general a superspace may be denoted by
${\mathbf{R}}^{p|q}$,  where   $p$, $q$ are the number of  real
commuting (bosonic) and anti-commuting (fermionic)
variables respectively. A supermatrix which takes
${\mathbf{R}}^{p|q}\rightarrow{\mathbf{R}}^{p|q}$
may be represented by a $(p+q)\times (p+q)$ matrix,
$M$, of the form
\begin{equation}
M=\left(\begin{array}{cc}
a&b\\
c& d\end{array}\right)
\end{equation}
where $a,d$ are $p\times p$, $q\times q$ matrices of
Grassmanian even
or bosonic variables and
$b,c$ are $p\times q$, $q\times p$ matrices of  Grassmanian odd or
fermionic variables respectively.\newline
The inverse  of $M$ can be expressed as
\begin{equation}
M^{-1}=\left(\begin{array}{cc}
(a-bd^{-1}c)^{-1}&-a^{-1}b(d-ca^{-1}b)^{-1}\\
-d^{-1}c(a-bd^{-1}c)^{-1}&(d-ca^{-1}b)^{-1}
\end{array}\right)
\end{equation}
where we may write
\begin{equation}
(a-bd^{-1}c)^{-1}=a^{-1}+\sum_{n=1}^{\infty}\,(a^{-1}bd^{-1}c)^{n}a^{-1}
\end{equation}
Note that due to the fermionic property of $b,c$, the power series
terminates at $n\leq pq+1$. \newline
%%%Hence, $M^{-1}$ exists if $a^{-1}$
%%%and $d^{-1}$ exist.\newline
The supertrace and the superdeterminant of $M$ are defined as
\begin{eqnarray}
&\mbox{str}\,M=\mbox{tr}\,a-\mbox{tr}\,d&\\
{}\nonumber\\
&\mbox{sdet}\,M=\det(a-bd^{-1}c)/\det d=\det a/\det (d-ca^{-1}b)&
\label{sdetdef}
\end{eqnarray}
The last equality comes from
\begin{equation}
\det(1-a^{-1}bd^{-1}c)=\mbox{det}{}^{-1} (1-d^{-1}ca^{-1}b)
\end{equation}
which may be shown using
\begin{equation}
\det
(1-a)=\mbox{exp}\left(
-\sum_{n=1}^{\infty}\,
\frac{1}{n}\mbox{tr}\,a^{n}\right)
\end{equation}
and observing
\begin{equation}
\mbox{tr}\,(a^{-1}bd^{-1}c)^{n}=-\mbox{tr}\,(d^{-1}ca^{-1}b)^{n}
\end{equation}
From eq.(\ref{sdetdef}) we note that
$\mbox{sdet}\,M\neq 0$ implies the existence of $M^{-1}$.
Thus the set of supermatrices for $\mbox{sdet}\,M\neq 0$ forms the
supergroup, $\mbox{Gl}(p|q)$. If  $\mbox{sdet}\,M=1$ then \newline
$M\in\mbox{Sl}(p|q)$. \newline
The supertrace and the superdeterminant have the properties
\begin{eqnarray}
&\mbox{str}\,(M_{1}M_{2})=\mbox{str}\,(M_{2}M_{1})&\\
{}\nonumber\\
&\mbox{sdet}\,(M_{1}M_{2})=\mbox{sdet}\,M_{1}\,\mbox{sdet}\,M_{2}&
\end{eqnarray}
We may define the transpose of the supermatrix, $M$,  either as
\begin{equation}
M^{t}=\left(\begin{array}{cc}
a^{t}&c^{t}\\
-b^{t}& d^{t}\end{array}\right)
\end{equation}
or as
\begin{equation}
M^{t^{\prime}}=\left(\begin{array}{cc}
a^{t}&-c^{t}\\
b^{t}& d^{t}\end{array}\right)
\end{equation}
where $a^{t},b^{t},c^{t},d^{t}$ are the ordinary transposes of
$a,b,c,d$ respectively.\newline
We note that
\begin{eqnarray}
&(M_{1}M_{2})^{t}=M_{2}^{t}M_{1}^{t}~~~~~~~~
(M_{1}M_{2})^{t^{\prime}}=M_{2}^{t^{\prime}}M_{1}^{t^{\prime}}&\\
{}\nonumber\\
&(M^{t})^{t^{\prime}}=(M^{t^{\prime}})^{t}=M&
\end{eqnarray}

\newpage
%%%%%%%%%%%%%%%%%%%%%%%%%%%%%%%%%%%%%%%%%%%%%%%%%%%%%%%%%%%%%%%%%%%%%%%%%%%%%%%%%%%%%%%%%%%%%%%%%%%%%%%%%%%%%%%%%%%%%%%%%%%%%%%%%%%%%%%%%%%%%%%%%%%%%%%%%%%%%%%%%%%%%%%%%%%%%%%%%%%%%%%%%%%%%%%%%%%%%%%%%%%%%%%%%%%%%%%%%%%%%%%%%%%%%%%%%%%%%
%%%%%%%%%%%%%%%%%%%%%%%%%%%%%%%%%%%%%%%%%%%%%%%%%%%%%%%%%%%%%%%%%%%%%%%%%%%%%%%%%%%%%%%%%%%%%%%%%%%%%%%%%%%%%%%%%%%%%%%%%%%%%%%%%%%%%%%%%%%%%%%%%%%%%%%%%%%%%%%%%%%%%%%%%%%%%%%%%%%%%%%%%%%%%%%%%%%%%%%%%%%%%%%%%%%%%%%%%%%%%%%%%%%%%%%%%%%%%
%%%%%%%%%%%%%%%%%%%%%%%%%%%%%%%%%%%%%%%%%%%%%%%%%%%%%%%%%%%%%%%%%%%%%%%%%%%%%%%%%%%%%%%%%%%%%%%%%%%%%%%%%%%%%%%%%%%%%%%%%%%%%%%%%%%%%%%%%%%%%%%%%%%%%%%%%%%%%%%%%%%%%%%%%%%%%%%%%%%%%%%%%%%%%%%%%%%%%%%%%%%%%%%%%%%%%%%%%%%%%%%%%%%%%%%%%%%%%
%%%%%%%%%%%%%%%%%%%%%%%%%%%%%%%%%%%%%%%%%%%%%%%%%%%%%%%%%%%%%%%%%%%%%%%%%%%%%%%%%%%%%%%%%%%%%%%%%%%%%%%%%%%%%%%%%%%%%%%%%%%%%%%%%%%%%%%%%%%%%%%%%%%%%%%%%%%%%%%%%%%%%%%%%%%%%%%%%%%%%%%%%%%%%%%%%%%%%%%%%%%%%%%%%%%%%%%%%%%%%%%%%%%%%%%%%%%%%
\section{Super Yang-Mills}
\subsection{$(3+1)D$  $\N=1$ super Yang-Mills}
In four-dimensional Minkowskian spacetime of the metric, $\eta=\diag(-+++)$, the $4\times 4$  gamma matrices satisfy with $\mu=0,1,2,3$,
\be
\ba{lll}
\Gamma^{\mu}{}^{\dagger}=\Gamma_{\mu}=-A\Gamma^{\mu}A^{\dagger}\,,
~~~~&~~~~A=\Gamma^{t}=-A^{\dagger}\,,\\
{}&{}\\
\Gamma^{\mu}{}^{\ast}=+B\Gamma^{\mu}B^{\dagger}\,,
~~~~&~~~~B^{T}=B\,,~~~~&~~~~B^{\dagger}=B^{-1}\,,\\
{}&{}\\
\Gamma^{\mu}{}^{T}=-C\Gamma^{\mu}C^{\dagger}\,,
~~~~&~~~~C=-C^{T}=B\Gamma^{t}\,,~~~~&~~~~C^{\dagger}=C^{-1}\,.
\ea
\ee
The Majorana spinor, $\psi$ satisfies then
\be
\ba{lll}
\bar{\psi}=\psi^{\dagger}\Gamma^{t}=\psi^{T}C
~~~&~~~\Longleftrightarrow~~~&~~~\psi^{\ast}=B\psi\,.
\ea
\ee
The four-dimensional super Yang-Mills Lagrangian reads
\begin{equation}
\dis{\L_{4D}=\tr\left(-\textstyle{\frac{1}{4}}F_{\mu\nu}F^{\mu\nu}
-i\half\bar{\psi}\Gamma^{\mu}D_{\mu}\psi\right)\,.}
\end{equation}
The supersymmetry transformations are
\be
\ba{ll}
\delta A_{\mu}=i\bar{\varepsilon}\Gamma_{\mu}\psi=-i\bar{\psi}\Gamma_{\mu}\varepsilon\,,~~~~&~~~~
\delta\psi=-\textstyle{\frac{1}{2}}F_{\mu\nu}\Gamma^{\mu\nu}\varepsilon\,.
\ea
\ee

%%%%%%%%%%%%%%%%%%%%%%%%%%%%%%%%%%%%%%%%%%%%%%%%%%%%%%%%%%%%%%%%%%%%%%%%%%%%%%%%%%%%%%%%%%%%%
%%%%%%%%%%%%%%%%%%%%%%%%%%%%%%%%%%%%%%%%%%%%%%%%%%%%%%%%%%%%%%%%%%%%%%%%%%%%%%%%%%%%%%%%%%%%%
%%%%%%%%%%%%%%%%%%%%%%%%%%%%%%%%%%%%%%%%%%%%%%%%%%%%%%%%%%%%%%%%%%%%%%%%%%%%%%%%%%%%%%%%%%%%%

\subsection{ $(5+1)D$  $(1,0)$ super Yang-Mills}
In six-dimensional Minkowskian spacetime of the metric, $\eta=\diag(-+++++)$, the $8\times 8$  gamma matrices satisfy with $M=0,1,2,3,4,5$,
\be
\ba{lll}
\Gamma^{M}{}^{\dagger}=\Gamma_{M}=A\Gamma^{M}A^{\dagger}\,,
~~~~&~~~~A:=\Gamma^{12345}=A^{\dagger}=A^{-1}\,,\\
{}&{}\\
\Gamma^{M}{}^{T}=C\Gamma^{M}C^{\dagger}\,,
~~~~&~~~~C^{T}=-C\,,~~~~&~~~~C^{\dagger}=C^{-1}\,,\\
{}&{}\\
\Gamma^{M}{}^{\ast}=B\Gamma^{M}B^{\dagger}\,,
~~~~&~~~~B=CA=-B^{T}\,,~~~~&~~~~B^{\dagger}=B^{-1}\,.
\ea
\ee
The gamma ``seven" is given by
$\Gamma^{(7)}=\Gamma^{012345}$ to satisfy $\Gamma^{(7)}=\Gamma^{(7)\dagger}=\Gamma^{(7)-1}$ and
\be
\Gamma^{LMN}=\textstyle{\frac{1}{6}}\epsilon^{LMNPQR}\,\Gamma_{PQR}\Gamma^{(7)}\,,
\label{self-dual}
\ee
where $\epsilon^{012345}=+1$.\\

The $\mbox{su}(2)$ Majorana-Weyl spinor, $\psi_{i}$, $i=1,2$, satisfies then
\be
\ba{ll}
\Gamma^{(7)}\psi_{i}=+\psi_{i}\,,~~~~~~~\bar{\psi}^{i}\Gamma^{(7)}=-\bar{\psi}^{i}
~~~~&~~~~:~~\mbox{chiral}\,,\\
{}\\
\bar{\psi}^{i}=(\psi_{i})^{\dagger}A
=\epsilon^{ij}(\psi_{j})^{T}C~~~~&~~~~:~~\mbox{su(2)~Majorana}\,,
\ea
\label{su2MW}
\ee
where $\epsilon^{ij}$ is the usual  $2\times 2$ skew-symmetric unimodular matrix.  It is worth to note that
$\bar{\psi}^{i}\Gamma^{M_{1}M_{2}\cdots M_{{2n}}}\rho_{i}=0$ and
\be
\tr(i\bar{\psi}^{i}\Gamma^{M_{1}M_{2}\cdots M_{{2n+1}}}\rho_{i})=\left[\tr(i\bar{\psi}^{i}\Gamma^{M_{1}M_{2}\cdots M_{{2n+1}}}\rho_{i})\right]^{\dagger}=-(-1)^{n}\tr(i\bar{\rho}^{i}\Gamma^{M_{1}M_{2}\cdots M_{{2n+1}}}\psi_{i})\,,
\label{reality}
\ee
where $\psi_{i}$,  $\rho_{i}$ are two arbitrary   Lie algebra valued $\mbox{su}(2)$ Majorana-Weyl spinors.\\

The  six-dimensional super Yang-Mills Lagrangian  reads
\begin{equation}
\dis{\L_{6D}=\tr\left(-\textstyle{\frac{1}{4}}F_{LM}F^{LM}
-i\half\bar{\psi}^{i}\Gamma^{L}D_{L}\psi_{i}\right)\,,}
\end{equation}
where all the fields are in the adjoint representation of the gauge group such that, with the Hermitian Lie algebra valued gauge field, $A_{M}$,
\be
\ba{ll}
D_{L}\psi_{i}=\partial_{L}\psi_{i}-i[A_{L},\psi_{i}]\,,~~~~&~~~~ F_{LM}=\partial_{L}A_{M}-\partial_{M}A_{L}-i[A_{L},A_{M}]\,.
\ea
\ee
From (\ref{reality}) the action  is real valued. \\

The supersymmetry transformations are given by with a $\mbox{su}(2)$ Majorana-Weyl supersymmetry parameter, $\varepsilon_{i}$,
\be
\ba{ll}
\delta A_{M}=+i\bar{\varepsilon}^{i}\Gamma_{M}\psi_{i}=
-i\bar{\psi}^{i}\Gamma_{M}\varepsilon_{i}\,,~~~~&~~~~
\delta\psi_{i}=-\textstyle{\frac{1}{2}}F_{MN}\Gamma^{MN}\varepsilon_{i}\,,
\ea
\label{susy0}
\ee
so that, in particular, $\delta\bar{\psi}^{i}=+\textstyle{\frac{1}{2}}F_{MN}\bar{\varepsilon}^{i}\Gamma^{MN}$. The crucial Fierz identity for the supersymmetry invariance  is with the chiral projection matrix, $P:=\half (1+\Gamma^{(7)})$,
\be
\left(\Gamma^{L}P\right)_{\alpha\beta}
\left(\Gamma_{L}P\right)_{\gamma\delta}+
\left(\Gamma^{L}P\right)_{\gamma\beta}
\left(\Gamma_{L}P\right)_{\alpha\delta}=0\,,
\label{Fierz}
\ee
which ensures the vanishing of the terms cubic in $\psi_{i}$,
\be
\tr\left(\bar{\psi}^{i}\Gamma^{L}
[\delta A_{L},\,\psi_{i}]\right)=
\tr\left(\bar{\psi}^{i}\Gamma^{L}
[\,i\bar{\varepsilon}^{j}\Gamma_{L}\psi_{j}\,,\,\psi_{i}]\right)=0\,.
\label{Fierz2}
\ee
~\\

The equations of motion are
\be
\ba{ll}
D_{L}F^{LM}+\bar{\psi}^{i}\Gamma^{M}\psi_{i}=0\,,~~~~&~~~~\Gamma^{M}D_{M}\psi_{i}=0\,.
\ea
\ee
\newpage

%%%%%%%%%%%%%%%%%%%%%%%%%%%%%%%%%%%%%%%%%%%%%%%%%%%%%%%%%%%%%%%%%%%%%%%%%%%%%%%%%%%%%%%%%%%%%
%%%%%%%%%%%%%%%%%%%%%%%%%%%%%%%%%%%%%%%%%%%%%%%%%%%%%%%%%%%%%%%%%%%%%%%%%%%%%%%%%%%%%%%%%%%%%
%%%%%%%%%%%%%%%%%%%%%%%%%%%%%%%%%%%%%%%%%%%%%%%%%%%%%%%%%%%%%%%%%%%%%%%%%%%%%%%%%%%%%%%%%%%%%

\subsection{ $6D$ super Yang-Mills in the spacetime  of arbitrary signature}
With
\be
\ba{ll}
\dis{\left(\Gamma^{M}\right)^{T}=\pm\C_{\pm}\Gamma^{M}\C_{\pm}^{-1}\,,}
~~~~&~~~~\C_{\pm}^{T}=\mp \C_{\pm}\,,
\ea
\ee
we have
\be
\dis{\left(\C_{\pm}\Gamma^{M}\right)^{T}=-\C_{\pm}\Gamma^{M}\,.}
\ee
We introduce a pair of Weyl spinors of the same chirality,
\be
\ba{lll}
\left(\psi_{1},\psi_{2}\right)\,, ~~~&~~~\Gamma^{(7)}\psi_{i}=s\psi_{i}\,,~~&~~s^{2}=1\,,
\ea
\ee
and define the charge conjugate spinor by
\be
\dis{\bar{\psi}_{c}^{i}:=\epsilon^{-1}{}^{ij}\psi_{j}^{T}\C_{\pm}\,.}
\ee

The super Yang-Mills Lagrangian reads
\be
\dis{\L_{6D}=\tr\left(\textstyle{\frac{1}{4}}F_{MN}F^{MN}+\textstyle{\frac{1}{2}}
\bar{\psi}^{i}_{c}\Gamma^{M}D_{M}\psi_{i}\right)\,,}
\ee
and the supersymmetry transformations are given by
\be
\ba{l}
\delta A_{M}=\bar{\varepsilon}_{c}^{i}\Gamma_{M}\psi_{i}=
-\bar{\psi}_{c}^{i}\Gamma_{M}\varepsilon_{i}\,,\\
{}\\
\delta\psi_{i}=-\textstyle{\frac{1}{2}}F_{MN}\Gamma^{MN}\varepsilon_{i}\,,
\ea
\ee
so that, in particular, $\delta\bar{\psi}_{c}^{i}=+\textstyle{\frac{1}{2}}F_{MN}\bar{\varepsilon}_{c}^{i}\Gamma^{MN}$.
The Lagrangian transforms as, from (\ref{symag}),
\be
\dis{\delta \L_{6D}=\partial_{M}\tr\left(F^{MN}\delta A_{N}-\textstyle{\frac{1}{2}}
\bar{\psi}^{i}_{c}\Gamma^{M}\delta\psi_{i}\right)\,.}
\ee

Only if $B_{\pm}^{\ast}B_{\pm}=-1$, as in the Minkowskian signature,  one can impose the pseudo-Majorana condition,
\be
\bar{\psi}_{c}^{i}=\bar{\psi}_{D}^{i}:=(\psi_{i})^{\dagger}A\,.
\ee

%%%%%%%%%%%%%%%%%%%%%%%%%%%%%%%%%%%%%%%%%%%%%%%%%%%%%%%%%%%%%%%%%%%%%%%%%%%%%%%%%%%%%%%%%%%%%
%%%%%%%%%%%%%%%%%%%%%%%%%%%%%%%%%%%%%%%%%%%%%%%%%%%%%%%%%%%%%%%%%%%%%%%%%%%%%%%%%%%%%%%%%%%%%
%%%%%%%%%%%%%%%%%%%%%%%%%%%%%%%%%%%%%%%%%%%%%%%%%%%%%%%%%%%%%%%%%%%%%%%%%%%%%%%%%%%%%%%%%%%%%
\newpage
\subsection{$(9+1)D$  SYM, its reduction, and  $4D$ superconformal symmetry}
\begin{itemize}
\item {\bf Conventions for $(9+1)D$ gamma matrices}\\
Spacetime signature : $\eta=\mbox{diag}(-++\cdots+)$, mostly plus signature.\\
$32\times 32$ Gamma matrices:\\
i) Hermitian conjugate,
\be
\ba{l}
(\Gamma^{M})^{\dagger}
=\Gamma_{M}=-\Gamma^{0}\Gamma^{M}\Gamma_{0}=\A\Gamma^{M}\A^{\dagger}\,,\\
{}\\
\A=\Gamma^{12\cdots 9}=\A^{\dagger}=\A^{-1}\,,\\
{}\\
(\A\Gamma^{M_{1}M_{2}\cdots M_{n}})^{\dagger}
=(-1)^{\frac{1}{2}n(n-1)}\A\Gamma^{M_{1}M_{2}\cdots M_{n}}\,,
\ea
\ee
{}\\
ii) Complex conjugate,
\be
\ba{l}
(\Gamma^{M})^{\ast}
=\pm \B_{\pm}\Gamma^{M}\B_{\pm}^{\dagger}\,,\\
{}\\
\B_{\pm}=\B_{\pm}^{T}=(\B_{\pm}^{\dagger})^{-1}\,,
\ea
\ee
{}\\
iii)  Transpose,
\be
\ba{l}
(\Gamma^{M})^{T}
=\pm \C_{\pm}\Gamma^{M}\C_{\pm}^{\dagger}\,,\\
{}\\
\C_{\pm}=\B_{\pm}^{T}\A=\pm\C_{\pm}^{\,T}=(\C^{\dagger}_{\pm})^{-1}\,,\\
{}\\
(\C_{+}\Gamma^{M_{1}M_{2}\cdots M_{n}})^{T}
=(-1)^{\frac{1}{2}n(n-1)}\C_{+}\Gamma^{M_{1}M_{2}\cdots M_{n}}\,.
\ea
\ee

Let the spinorial indices be located as
\be
\ba{llll}
(\Gamma^{M})^{\alpha}{}_{\beta}\,,~~&~~(\A)^{\alpha}{}_{\beta}\,,~~&~~
(\B_{\pm})_{\alpha\beta}=(\B_{\pm})_{\beta\alpha}\,,~~&~~(\C_{\pm})_{\alpha\beta}=
\pm(\C_{\pm})_{\beta\alpha}\,.
\ea
\ee

Define
\be
\Gamma^{(10)}=\Gamma^{012\cdots 9}=(\Gamma^{(10)})^{\dagger}=(\Gamma^{(10)})^{-1}=
-\C_{+}^{\dagger}(\Gamma^{(10)})^{T}\C_{+}\,.\label{g11}
\ee

The crucial identity for the super Yang-Mills action is
\be
(\C_{+}\Gamma^{M}\Gamma_{\pm})_{(\alpha\beta}
(\C_{+}\Gamma_{M}\Gamma_{\pm})_{\gamma)\delta}
=0
%%+
%(\C_{+}^{-1}\Gamma^{M}\Gamma_{\pm})_{\gamma\alpha}
%(\C_{+}^{-1}\Gamma_{M}\Gamma_{\pm})_{\beta\delta}+
%(\C_{+}^{-1}\Gamma^{M}\Gamma_{\pm})_{\beta\gamma}
%(\C_{+}^{-1}\Gamma_{M}\Gamma_{\pm})_{\alpha\delta}=0\,,
\label{crucial}
\ee
where $\Gamma_{\pm}=\half(1\pm\Gamma^{(10)})$ is either the chiral or the anti-chiral projector,  and $\alpha,\beta,\gamma$  are symmetrized.
Note also the symmetric property,
$(\C_{+}\Gamma^{M}\Gamma_{\pm})^{T}=\C_{+}\Gamma^{M}\Gamma_{\pm}$.\\

For spinors we set
\be
\bar{\psi}=\psi^{\dagger}\A\,.
\ee

Majorana-Weyl Spinor, $\psi$, satisfies
\begin{eqnarray}
{}&{}&\Gamma^{(10)}\psi=+\psi~~~~~~~:~~\mbox{Weyl~condition}\,,\\
{}&{}&{}\nonumber\\
{}&{}&\psi^{\ast}=\B_{+}\psi~~~~~~~~~~:~~\mbox{Majorana~condition}\,,
\end{eqnarray}
or equivalently,
\be
\ba{l}
\bar{\psi}\Gamma^{(10)}=-\bar{\psi}~~:~\mbox{opposite~chirality}\,,\\
{}\\
\bar{\psi}=\psi^{T}\C_{+}\,.
\ea
\ee
Hence for the fermionic  Majorana-Weyl spinors,
\be
\bar{\psi}_{1}\Gamma^{M_{1}M_{2}\cdots M_{2n}}\psi_{2}=0\,,
\label{Ap1}
\ee
and\footnote{When the spinor is Lie algebra valued, Eq.(\ref{Ap2}) does not hold in general.}
\be
\bar{\psi}_{1}\Gamma^{M_{1}M_{2}\cdots M_{2n+1}}\psi_{2}=
(-1)^{n+1}\bar{\psi}_{2}\Gamma^{M_{1}M_{2}\cdots M_{2n+1}}\psi_{1}=
-(\bar{\psi}_{1}\Gamma^{M_{1}M_{2}\cdots M_{2n+1}}\psi_{2})^{\dagger}~~:~\mbox{imaginary}\,.
\label{Ap2}
\ee
~~\\

We can further set
\be
\ba{lll}
\Gamma^{M}=\left(\ba{cc}\,0\,&\,\tilde{\gamma}^{M}\\
\,\gamma^{M}&\,0\,\ea\right)\,,~~~&~~~\gamma^{M}\tilde{\gamma}^{N}+
\gamma^{N}\tilde{\gamma}^{M}=2\eta^{MN}\,,~~~&~~~\eta=\mbox{diag}(-+++\cdots +)\,.
\ea
\ee
Namely,  $(\gamma^{M},\tilde{\gamma}^{N})$ are the  real  $16\times 16$ matrices  appearing in the off block-diagonal parts of the $32\times 32$ gamma matrices,

satisfying\footnote{From $\tilde{\gamma}^{M}=(\gamma_{M})^{-1}$ it also follows that $\tilde{\gamma}^{M}\gamma^{N}+\tilde{\gamma}^{N}\gamma^{M}=2\eta^{MN}$. One may further impose the symmetric property, $(\gamma^{M})^{T}=\gamma^{M}$, but it is not necessary in our paper.}
\be
\ba{ll}
(\gamma^{M})^{\ast}=\gamma^{M}\,,~~~~~&~~~~~
(\gamma^{M})^{T}=\tilde{\gamma}^{0}\gamma^{M}\tilde{\gamma}^{0}=\tilde{\gamma}_{M}\,,\\
{}&{}\\
\tilde{\gamma}^{0}\gamma^{1}\tilde{\gamma}^{2}\cdots\gamma^{9}=+1\,,~~~~~&~~~~~
{\gamma}^{0}\tilde{\gamma}^{1}\gamma^{2}\cdots\tilde{\gamma}^{9}=-1\,.
\ea
\ee

%%%%%%%%%%%%%%%%%%%%%%%%%%%%%%%%%%%%%%%%%%%%%%%%%%%%%%%%%%%%%%%%%%%%%%%%%%%%%%%%%%%%%%%%%%%%%%%%
%%%%%%%%%%%%%%%%%%%%%%%%%%%%%%%%%%%%%%%%%%%%%%%%%%%%%%%%%%%%%%%%%%%%%%%%%%%%
%%%%%%%%%%%%%%%%%%%%%%%%%%%%%%%%%%%%%%%%%%%%%%%
\item {\bf Lagrangian.}\\
Let the gauge group be $\su(N)$ or $\mbox{u}(N)$.\\
Lie algebra valued fields,
\be
\ba{lll}
A_{M}=A_{M}^{p}T_{p}\,,~~&~~\Psi=\Psi^{p}T_{p}\,,~~&~~(T_{p})^{\dagger}=T_{p}\,.
\ea
\ee
Field strength and the covariant derivative are
\be
\ba{ll}
F_{MN}=\partial_{M}A_{N}-\partial_{N}A_{M}-i[A_{M},A_{N}]\,,~~&~~D_{M}\Psi=
\partial_{M}\Psi-i[A_{M},\Psi]\,.
\ea
\ee
Bianchi identity reads
\be
D_{L}F_{MN}+D_{M}F_{NL}+D_{N}F_{LM}=0\,.
\label{Bianchi}
\ee

The gauge symmetry is given by, for $g^{\dagger}=g^{-1}$,
\be
\ba{lll}
A_{M}\rightarrow gA_{M}g^{-1}+ig\partial_{M}g^{-1}\,,~~&~~
F_{MN}\rightarrow gF_{MN}g^{-1}\,,~~&~~\Psi\rightarrow g\Psi g^{-1}\,.
\ea
\ee
{}\\

The Lagrangian of $10D$ super Yang-Mills theory reads
\be
\ba{ll}
\L&=\tr\left[-\textstyle{\frac{1}{4}}F_{MN}F^{MN}-i\half\bar{\Psi}\Gamma^{M}D_{M}\Psi
\right]\\
{}&{}\\
{}&=\tr\left[-\textstyle{\frac{1}{4}}F_{MN}F^{MN}-i\half\bar{\psi}\gamma^{M}D_{M}\psi\right]\,,
\ea
\ee
where $\Psi\equiv(\psi~0)^{T}$ and  $\psi^{\alpha}$ is a  sixteen component spinor and $\bar{\psi}:=\psi^{T}\tilde{\gamma}^{0}$. \\

Under arbitrary infinitesimal transformations, $\delta A_{M}$, $\delta\Psi$,
\be
\delta\L=\tr\left[\left(D_{L}F^{LM}+\bar{\Psi}\Gamma^{M}\Psi\right)\delta A_{M}-i\bar{\Psi}\Gamma^{M}D_{M}\delta\Psi\right]
+\partial_{N}\tr\left[F^{MN}\delta A_{M}-i\half\delta\bar{\Psi}\Gamma^{N}\Psi\right]\,.
\label{infL}
\ee
~~\\
~~\\
~~\\

%%%%%%%%%%%%%%%%%%%%%%%%%%%%%%%%%%%%%%%%%%%%%%%%%%%%%%%%%%%%%%%%%%%%%%%%%%%%%%%%%%%%%%%%%%%%%%%%%%%%%%%%%%%%%%%%%%%%%%%%%%%%%
%%%%%%%%%%%%%%%%%%%%%%%%%%%%%%%%%%%%%%%%%%%%%%%%%%%%%%%%%%%%%%%%%%%%%%%%%%%%%%%%%%%%%%%%%%%%%%%%%%%%%%%%%%%%%%%%%%%%%%%%%%%%%
\item {\bf Summary of supersymmetry in $D\leq 10$.}\\
The ordinary supersymmetry and kinetic supersymmetry are given by
\be
\ba{ll}
\delta A_{M}=i\bar{\Psi}\Gamma_{M}\xi_{+}=-i\bar{\xi}_{+}\Gamma_{M}\Psi\,,~~&~~
\delta\Psi=\half F_{MN}\Gamma^{MN}\xi_{+}+\xi_{+}^{\prime}\I_N\,,
\ea
\ee
so that
\be
\delta\bar{\Psi}=-\half\bar{\xi}_{+}F_{MN}\Gamma^{MN}+\bar{\xi}_{+}^{\prime}\I_N\,,
\ee
where $\xi_{+}$ and $\xi_{+}^{\prime}$ are constant Majornana-Weyl spinors corresponding to the ordinary and kinetic supersymmetry parameters. $+$ denotes the chirality. The above is the symmetry of the $(9+1)D$ and also  any dimensionally reduced super Yang-Mills action.\newline
{}\\
In four-dimensions of either Minkowskian or Euclidean signature, the  supersymmetry gets enhanced to the superconformal symmetry as
\be
\ba{ll}
\delta A_{M}=i\bar{\Psi}\Gamma_{M}\E(x)=-i\bar{\E}(x)\Gamma_{M}\Psi\,,~~&~~
\delta\Psi=\half F_{MN}\Gamma^{MN}\E(x)-2\Phi_{a}\Gamma^{a}\xi_{-}+\xi_{+}^{\prime}\I_N\,,
\ea
\ee
where $m$ is for the four-dimensions and $a$ is for the rest. $\xi_{-}$ is a constant Majornana-Weyl spinor of the opposite chirality corresponding to the special superconformal symmetry parameter, and
\be
\E(x)=x^{m}\Gamma_{m}\xi_{-}+\xi_{+}\,.
\ee
~\\
In any case, the conserved supercurrent is of the universal form,
\be
J^{M}=-i\tr\left(\bar{\Psi}\Gamma^{M}\delta\Psi\right)=
+i\tr\left(\delta\bar{\Psi}\Gamma^{M}\Psi\right)\,.
\ee
~\\
In Appendix \ref{ProofSec}, we present the derivation.
\newpage
%%%%%%%%%%%%%%%%%%%%%%%%%%%%%%%%%%%%%%%%%%%%%%%%%%%%%%%%%%%%%%%%%%%%%%%%%%%%%%%%%%%%%%%%%%%%%%%%%%%%%%%%%%%%%%%%%%%%%%%%%%%%%
%%%%%%%%%%%%%%%%%%%%%%%%%%%%%%%%%%%%%%%%%%%%%%%%%%%%%%%%%%%%%%%%%%%%%%%%%%%%%%%%%%%%%%%%%%%%%%%%%%%%%%%%%%%%%%%%%%%%%%%%%%%%%
\item {\bf Superconformal symmetry in $4D$ of arbitrary signature.}\\
The 32 supersymmetries in $4D$ super Yang-Mills which consist of ordinary supersymmetry and special superconformal symmetry read
\be
\ba{l}
\delta A_{M}=i\bar{\Psi}\Gamma_{M}(1+x^{m}\Gamma_{m})\VE
=-i\bar{\VE}(1+x^{m}\Gamma_{m})\Gamma_{M}\Psi\,,\\
{}\\
\delta\Psi=\half(1+\Gamma^{(10)})\left[\half F_{MN}\Gamma^{MN}(1+x^{m}\Gamma_{m})-2\Phi_{a}\Gamma^{a}\right]\VE\,,\\
{}\\
\delta\bar{\Psi}=\bar{\VE}\left[-\half (1+x^{m}\Gamma_{m}) F_{MN}\Gamma^{MN}-2\Phi_{a}\Gamma^{a}\right]\half(1-\Gamma^{(10)})\,,
\ea
\label{32SUSY}
\ee
where $\VE$ is a 32 component Majorana spinor,
\be
\VE^{\ast}=\B_{+}\VE\,.
\ee
The chiral decomposition of  the spinor gives the ordinary supersymmetry and special superconformal symmetry,\footnote{Note also $\E(x)=\half(1+\Gamma^{(10)})(1+x^{m}\Gamma_{m})\VE\,.$}
\be
\ba{ll}
\VE=\xi_{+}+\xi_{-}\,,~~~~&~~~~
\xi_{\pm}=\half(1\pm\Gamma^{(10)})\VE\,.
\ea
\ee

The 32 component Majorana supercurrent is of the form,
\be
\ba{l}
J^{M}=+i\bar{\Q}^{M}\VE=-i\bar{\VE}\Q^{M}\,,\\
{}\\
\Q^{M}=\tr\left[\left(\half (1+x^{m}\Gamma_{m}) F_{KL}\Gamma^{KL}+2\Phi_{a}\Gamma^{a}\right)\Gamma^{M}\Psi\right]\,,\\
{}\\
\bar{\Q}^{M}=
\tr\left[\bar{\Psi}\Gamma^{M}\left(-\half F_{KL}\Gamma^{KL}(1+x^{m}\Gamma_{m})+2\Phi_{a}\Gamma^{a}\right)\right]
=(\Q^{M})^{\dagger}\A=(\Q^{M})^{T}\C_{+}\,.
\ea
\label{32supercharge}
\ee

The supercharge is given by
\be
\Q=\displaystyle{\int\! d^{3}x~\Q^{0}}\,.
\ee

\end{itemize}

%%%%%%%%%%%%%%%%%%%%%%%%%%%%%%%%%%%%%%%%%%%%%%%%%%%%%%%%%%%%%%%%%%%%%%%%%%%%
%%%%%%%%%

%%%%%%%%%%%%%%%%%%%%%%%%%%%%%%%%%%%%%%%%%%%%%%%%%%%%%%%%%%%%%%%%%%%%%%%%%%%%%%%%%%%%%%%%%%%%%%%%
%%%%%%%%%%%%%%%%%%%%%%%%%%%%%
\newpage
\appendix

\begin{center}
{\huge\bf{APPENDIX}}
\end{center}
\section{Proof of the Theorem~(\ref{lemma})\label{Proof}}
\textbf{Theorem~(\ref{lemma}):}\\
Any $N\times N$ matrix, $M$, satisfying $M^{2}=\lambda^{2}\I_N$, $\lambda\neq 0$, is diagonalizable.\\
{}\\
\textit{Proof}\\
Suppose for some $K$, $1\leq K\leq N$, we have found a basis,
\be
\{e_{a},~v_{r}~:~1\leq a\leq K ,~1\leq r\leq N-K\}
\ee
such that
\be
\ba{ll}
Me_{a}=\lambda_{a} e_{a}\,,~~~~&~~~\mbox{~for~} 1\leq a\leq K\,,\\
{}&{}\\
Mv_{r}=P^{s}{}_{r}v_{s}+h^{a}{}_{r}e_{a}\,,~~~~&~~~\mbox{~for~} K+1\leq r,s\leq N\,.
\ea
\ee
From $M^{2}=\lambda^{2}\I_N$,
\begin{equation}
\ba{l}
\lambda_{a}^{2}=\lambda^{2}\,,\\
{}\\
\lambda^{2}v_{r}=(P^{2})^{s}{}_{r}v_{s}+\left[(hP)^{a}{}_{r}+\lambda_{a}h^{a}{}_{r}\right]
e_{a}\,,
\ea
\end{equation}
and hence,
\be
\ba{l}
P^{2}=\lambda^{2}1_{\scriptscriptstyle (N-K)\times(N-K)}\,,\\
{}\\
(hP)^{a}{}_{r}+\lambda_{a}h^{a}{}_{r}=0\,.
\ea
\ee
The assumption holds for $K=1$ surely.
In order to construct $e_{K+1}$ we first consider an eigenvector of the $(N-K)\times(N-K)$ matrix, $P$,
\be
\ba{ll}
P^{r}{}_{s}c^{s}=\lambda_{K+1}c^{r}\,,~~~~&~~~~\lambda_{K+1}^{2}=\lambda^{2}\,,
\ea
\ee
and set
\be
\ba{l}
v=c^{r}v_{r}\,,~~~~~~~~~h^{a}=h^{a}{}_{r}c^{r}\,,\\
{}\\
Mv=\lambda_{K+1}v+h^{a}e_{a}\,.
\ea
\ee
Consequently
\be
\ba{ll}
(\lambda_{K+1}+\lambda_{a})h^{a}=0~~~~~~~&~~:~~~\mbox{not~~}a\mbox{~~sum\,,}
\ea
\ee
so that
\be
\ba{ll}
h^{a}=0~~~~&~~~~~~~\mbox{if~~~~~}\lambda_{K+1}+\lambda_{a}\neq 0\,.
\ea
\label{hn0}
\ee
We construct $e_{K+1}$,  with $K$ unknown coefficients, $d^{a}$, as
\be
e_{K+1}=v+d^{a}e_{a}\,.
\ee
From
\be
Me_{K+1}=\lambda_{K+1}e_{K+1}+\left[h^{a}+(\lambda_{a}-\lambda_{K+1})d^{a}\right]e_{a}\,,
\ee
we determine
\be
d^{a}=\left\{\ba{ll}
\displaystyle{\frac{h^{a}}{\,\lambda_{K+1}-\lambda_{a}}}~~~~&~~~~\mbox{if~~~}
\lambda_{K+1}\neq\lambda_{a}\,,\\
{}&{}\\
\mbox{any~number~}~~~~&~~~~\mbox{if~~~}
\lambda_{K+1}=\lambda_{a}\,.
\ea
\right.
\ee
From (\ref{hn0}) and $\lambda_{K+1}^{2}=\lambda_{a}^{2}=\lambda^{2}\neq 0$, we have
\be
Me_{K+1}=\lambda_{K+1}e_{K+1}\,.
\ee
This completes our proof.\\
{}\\
If we set a $N\times N$ invertible matrix, $S$, by
\be
\ba{lll}
(S)^{b}{}_{a}=(e_{a})^{b}\,,~~~~&~~~~Me_{a}=\lambda_{a}e_{a}\,,~~~~&~~~~1\leq a,b\leq N\,,
\ea
\ee
then
\be
S^{-1}MS=\diag(\lambda_{1},\lambda_{2},\cdots,\lambda_{N})\,.
\ee

\newpage

%%%%%%%%%%%%%%%%%%%%%%%%%%%%%%%%%%%%%%%%%%%%%%%%%%%%%%%%%%%%%%%%%%%%%%%%%%%%%%%%%%%%%%%%%%%%%%%%
%%%%%%%%%%%%%%%%%%%%%%%%%%%%%%%%%%%%%%%%%%%%%%%%%%%%%%%%%%%%%%%%%%%%%%%%%%%%
%%%%%%%%%%%%%%%%%%%%%%%%%%%%%%%%%%%%%%%%%%%%%%%
\section{Gamma matrices in 4,6,10,12 dimensions}

Our conventions are such that
\be
\ba{llll}
\hat{\gamma}^{m}~~&:~~&~~m=0,1,2,3~~&~~\mbox{for~~}1+3D\,,\\
{}&{}&{}\\
\gamma^{\mu}~~&:~~&~~\mu=1,2,\cdots,6~~&~~\mbox{for~~}2+4D\,,\\
{}&{}&{}\\
\gamma^{a}~~&:~~&~~a=7,8,\cdots,12~~&~~\mbox{for~~}0+6D\,,\\
{}&{}&{}\\
\Gamma^{M}~~&:~~&~~M=0,1,2,3,7,\cdots,12~~~~&~~\mbox{for~~}1+9D\,,\\
{}&{}&{}\\
{\bf{\Gamma}}^{\bf M}~~&:~~&~~{\bf M}=1,2,\cdots,12~~&~~\mbox{for~~}2+10D\,.
\ea
\ee
%%%%%%%%%%%%%%%%%%%%%%%%%%%%%%%%%%%%%%%%%%%%%%%%%%%%%%%%%%%%%%%%%%%%%%%%%%%%%%%%%%%%%%%%%%%%%%%%
%%%%%%%%%%%%%%%%%%%%%%%%%%%%%
\subsection{Four dimensions}
In Minkowskian four dimension of the metric, $\hat{\eta}=\diag(-+++)$, the gamma matrices satisfy
\be
\ba{ll}
\hat{\gamma}^{m}\hat{\gamma}^{n}+\hat{\gamma}^{n}\hat{\gamma}^{m}=2\hat{\eta}^{mn}\,,~~~~~&~~~~~
(\hat{\gamma}^{m})^{\dagger}=\hat{\gamma}_{m}\,,
\ea
\ee
where $m,n=0,1,2,3$.      The chiral matrix reads
\be
\hat{\gamma}^{(5)}=-i\hat{\gamma}^{0123}=(\hat{\gamma}^{(5)})^{-1}
=(\hat{\gamma}^{(5)})^{\dagger}\,.
\label{4Dc}
\ee

The three pairs of unitary matrices,  $\hat{A}_{\pm},\hat{B}_{\pm},\hat{C}_{\pm}$,
relate the hermitain conjugate, complex conjugate, and the transpose of the gamma matrices,
\be
\ba{ll}
\pm (\hat{\gamma}^{m})^{\dagger}=\hat{A}_{\pm}\hat{\gamma}^{m}\hat{A}_{\pm}^{\dagger}\,,~~~&~~
\hat{A}_{\pm}^{\dagger}\hat{A}_{\pm}=1\,,\\
{}\\
\pm (\hat{\gamma}^{m})^{\ast}=\hat{B}_{\pm}\hat{\gamma}^{m}\hat{B}_{\pm}^{\dagger}\,,~~~&~~
\hat{B}_{\pm}^{\dagger}\hat{B}_{\pm}=1\,,\\
{}\\
\pm (\hat{\gamma}^{m})^{T}=\hat{C}_{\pm}\hat{\gamma}^{m}\hat{C}_{\pm}^{\dagger}\,,~~~&~~
\hat{C}_{\pm}^{\dagger}\hat{C}_{\pm}=1\,.
\ea
\ee
Especially in Minkowskian four dimensions, they can be chosen further to  satisfy
\be
\ba{lll}
\hat{A}_{+}=-i\gamma^{123}\,,~~&~~
\hat{A}_{-}=-\hat{\gamma}^{0}\,,~~&~~\hat{A}_{-}=\hat{A}_{+}\hat{\gamma}^{(5)}\,,\\
{}&{}&{}\\
\hat{B}_{\pm}^{\ast}\hat{B}_{\pm}=\pm 1\,,~~&~~
\hat{B}_{\pm}^{T}=\pm\hat{B}_{\pm}\,,~~&~~
\hat{B}_{-}=\hat{B}_{+}\hat{\gamma}^{(5)}\,,\\
{}&{}&{}\\
{\hat{C}_{\pm}=\hat{B}_{+}^{T}\hat{A}_{\pm}=\hat{B}_{\pm}^{T}\hat{A}_{+}\,,}
~~&~~
\hat{C}_{\pm}^{T}=-\hat{C}_{\pm}\,,~~&~~
\hat{C}_{-}=\hat{C}_{+}\hat{\gamma}^{(5)}\,.
\ea
\label{4DABC}
\ee

%%%%%%%%%%%%%%%%%%%%%%%%%%%%%%%%%%%%%%%%%%%%%%%%%%%%%%%%%%%%%%%%%%%%%%%%%%%%%%%%%%%%%%%%%%%%%%%%
%%%%%%%%%%%%%%%%%%%%%%%%%%%%%
\subsection{Four to six dimensions}
Using the four dimensional gamma matrices above, one can construct the six dimensional gamma matrices in the off-block diagonal form,
\be
\ba{lll}
\gamma^{\mu}=\left(\begin{array}{cc}0&\rho^{\mu}\\ \bar{\rho}^{\mu}&0\end{array}\right)\,,~~&~~\mu=1,2,\cdots,6\,,~~&~~
\rho^{\mu}\bar{\rho}^{\nu}+\rho^{\nu}\bar{\rho}^{\mu}=2\eta^{\mu\nu}\,.
%%%
%%~~&~~\eta=\diag(--++++)\,.
%%%
\ea
\ee
With the relevant choice of the metric,
\be
\eta=\diag(--++++)\,,
\ee
we require $\bar{\rho}^{\mu}=(\rho_{\mu})^{\dagger}$ and set
%%%
%%\footnote{The $i$ factor is chosen for the later use as in (\ref{12later}).}
%%%
\be
\ba{ll}
\gamma^{1}=U\left(-i\tau_{2}\otimes 1\right)U^{\dagger}\,,~~~~&~~~~
\gamma^{m+2}=U\left(\tau_{1}\otimes \hat{\gamma}^{m}\right)U^{\dagger}\,,\\
{}&{}\\
\gamma^{6}=U\left(\tau_{1}\otimes\hat{\gamma}^{(5)}\right)U^{\dagger}\,,~~~~&~~~~
U=\left(\ba{cc}\hat{C}_{+}&0\\0&1\ea\right)\,.
\ea
\label{g7}
\ee
Explicitly with  (\ref{4Dc}), (\ref{4DABC})
\be
\ba{lll}
\rho^{1}=-\hat{C}_{+}\,,~~~~&~~~\rho^{m+2}=\hat{C}_{+}\hat{\gamma}^{m}\,,~~~~&~~~
\rho^{6}=\hat{C}_{-}\,,\\
{}&{}&{}\\
\bar{\rho}^{1}=+\hat{C}^{-1}_{+}\,,~~~~&~~~\bar{\rho}^{m+2}=\hat{\gamma}^{m}\hat{C}_{+}^{-1}\,,
~~~~&~~~\bar{\rho}^{6}=\hat{C}_{-}^{-1}\,.
\ea
\ee
Note
\be
\gamma^{(7)}=i\gamma^{1}\gamma^{2}\cdots\gamma^{6}=
\left(\ba{cc}1&0\\0&-1\ea\right)\,,
\ee
and especially the anti-symmetric property of the $4\times 4$ matrices,
\be
\ba{ll}
(\rho_{\mu})_{\alpha\beta}=-(\rho_{\mu})_{\beta\alpha}\,,~~~&~~~
(\bar{\rho}^{\mu})^{\alpha\beta}=
-\half\epsilon^{\alpha\beta\gamma\delta}(\rho^{\mu})_{\gamma\delta}\,.
\ea
\ee
The spinorial indices, $\alpha,\beta=1,2,3,4$, denote   the fundamental representation of  $\su(2,2)$.
It follows that $\{\rho^{\mu}\}$ and  $\{\bar{\rho}^{\mu}\}$ separately form
bases for the anti-symmetric $4\times 4$ matrices with the
completeness relation,
\be
\ba{ll}
\tr(\rho^{\mu}\bar{\rho}_{\nu})=4\delta^{\mu}_{~\nu}\,,~~~&~~~
(\rho^{\mu})_{\alpha\beta}(\bar{\rho}_{\mu})^{\gamma\delta}=2(
\delta_{\alpha}{}^{\delta}\delta_{\beta}{}^{\gamma}
-\delta_{\beta}{}^{\delta}\delta_{\alpha}{}^{\gamma})\,.
%%%%
%%%\\
%%{}&{}\\
%%\tr(\rho^{a}\bar{\rho}_{b})=4\delta^{a}_{~b}\,,~~~&~~~
%%(\rho^{a})_{\dalpha\dbeta}(\bar{\rho}_{a})^{\dgamma\ddelta}=2(
%%\delta_{\dalpha}{}^{\ddelta}\delta_{\dbeta}{}^{\dgamma}
%%-\delta_{\dbeta}{}^{\ddelta}\delta_{\dalpha}{}^{\dgamma})\,,
%%%%
\ea
\label{antisym}
\ee
On the other hand, Eq.(\ref{g7})
implies that\footnote{We put $\epsilon^{123456}=1$ and ``$[~]$" denotes
the standard anti-symmetrization with ``strength one".}
\be
\ba{ll}
\rho^{[\mu}\bar{\rho}^{\nu}\rho^{\lambda]}=
+i\textstyle{\frac{1}{6}}\epsilon^{\mu\nu\lambda\sigma\tau\kappa}
{\rho}_{[\sigma}\bar{\rho}_{\tau}{\rho}_{\kappa]}\,,~~&~~
\bar{\rho}^{[\mu}{\rho}^{\nu}\bar{\rho}^{\lambda]}=
-i\textstyle{\frac{1}{6}}\epsilon^{\mu\nu\lambda\sigma\tau\kappa}
\bar{\rho}_{[\sigma}{\rho}_{\tau}\bar{\rho}_{\kappa]}\,,
\ea
\label{iden}
\ee
so each of the sets
$\rho^{[\mu}\bar{\rho}^{\nu}\rho^{\lambda]}\equiv\rho^{\mu\nu\lambda}$ or
$\bar{\rho}^{[\mu}{\rho}^{\nu}\bar{\rho}^{\lambda]}\equiv\bar{\rho}^{\mu\nu\lambda}$
has only 10 independent components and forms a basis for
symmetric $4\times 4$ matrices,
\be
\ba{l}
\tr(\rho^{\mu\nu\lambda}
\bar{\rho}_{\sigma\tau\kappa})
=-i4\,\epsilon^{\mu\nu\lambda}{}_{\sigma\tau\kappa}
-24\delta^{[\mu}_{~\sigma}\delta^{\nu}_{~\tau}\delta^{\lambda]}_{~\kappa}\,,\\
{}\\
(\rho^{\mu\nu\lambda})_{\alpha\beta}
(\bar{\rho}_{\mu\nu\lambda})^{\gamma\delta}=-24(
\delta_{\alpha}{}^{\gamma}\delta_{\beta}{}^{\delta}
+\delta_{\beta}{}^{\gamma}\delta_{\alpha}{}^{\delta})\,.

%%%
%%\tr(\rho^{[a}\bar{b}^{\nu}\rho^{c]}\bar{\rho}_{[d}{\rho}_{e}\bar{\rho}_{f]})
%%=-4i\epsilon^{abc}{}_{def}
%%-24\delta^{[a}_{~d}\delta^{b}_{~e}\delta^{c]}_{~f}\,,\\
%%{}\\
%%(\rho^{[a}\bar{\rho}^{b}\rho^{c]})_{\dalpha\dbeta}
%%(\bar{\rho}_{[a}{\rho}_{b}\bar{\rho}_{c]})^{\dgamma\ddelta}=-24(
%%\delta_{\dalpha}{}^{\dgamma}\delta_{\dbeta}{}^{\ddelta}
%%+\delta_{\dbeta}{}^{\dgamma}\delta_{\dalpha}{}^{\ddelta})\,.
%%%%
\ea
\label{sym}
\ee
Finally,   $\{\rho^{\mu\nu}\equiv\frac{1}{2}({\rho}^{\mu}\bar{\rho}^{\nu}-{\rho}^{\nu}\bar{\rho}^{\mu})\}$  or $\{\bar{\rho}^{\mu\nu}\equiv\frac{1}{2}(\bar{\rho}^{\mu}{\rho}^{\nu}-\bar{\rho}^{\nu}{\rho}^{\mu})\}$  forms an orthonormal basis for the  general $4\times 4$ traceless matrices,
\begin{equation}
\ba{ll}
\tr(\rho^{\mu\nu}\rho_{\lambda\kappa})=4(\delta^{\mu}{}_{\kappa}\delta^{\nu}{}_{\lambda}
-\delta^{\nu}{}_{\kappa}\delta^{\mu}{}_{\lambda})\,,~&~
%%%
%% \label{6dorthonormal}
%%%
\textstyle{-\frac{1}{8}}(\rho^{\mu\nu})_{\alpha}{}^{\beta}(\rho_{\mu\nu})_{\gamma}{}^{\delta}+
\textstyle{\frac{1}{4}}\delta_{\alpha}{}^{\beta}\delta_{\gamma}{}^{\delta}
=\delta_{\alpha}{}^{\delta}\delta_{\gamma}{}^{\beta}\,,
\ea
\label{6dfierz}
\end{equation}
satisfying
\be
(\bar{\rho}^{\mu\nu})^{\alpha}{}_{\beta}=-(\rho^{\mu\nu})_{\beta}{}^{\alpha}\,.\label{rbr}
\ee

%%%%%%%%%%%%%%%%%%%%%%%%%%%%%%%%%%%%%%%%%%%%%%%%%%%%%%%%%%%%%%%%%%%%%%%%%%%%%%%%%%%%%%%%%%%%%%%%
%%%%%%%%%%%%%%%%%%%%%%%%%%%%%
\subsection{Six dimensions}
%%
%In order to make the $\mbox{SO}(2,4)\times\mbox{SO}(6)$ isometry of
%$AdS_{5}\times S^{5}$ geometry manifest, it is convenient to employ
%the twelve dimensional gamma matrices of spacetime signature
%$(--++++++++++)$, and write them in terms of two sets of
%six dimensional gamma matrices, $\{\gamma^{\mu}\}$, $\{\gamma^{a}\}$,
%%%
%\begin{equation}
%\begin{array}{ll}
%\Gamma^{\mu}=\gamma^{\mu}\otimes\gamma^{(7)}&~~~\mbox{for~}~\mu=1,2,3,4,5,6\\
%{}&{}\\
%\Gamma^{a}\,=\,1\,\otimes\gamma^{a}&~~~\mbox{for~}~a=7,8,9,10,11,12\,.
%\end{array}
%\end{equation}
%
The result above can be straightforwardly generalized to other signatures in six dimensions.
In Euclidean six dimensions,  gamma matrices satisfy
\be
\gamma^{a}\gamma^{b}+\gamma^{b}\gamma^{a}=2\delta^{ab}\,,
\ee
where we set $a,b$ run from 7 to 12,  instead of 1 to 6,  as the latter have been reserved for $\so(2,4)$. With the choice,
\be
\gamma^{(7)}=i\gamma^{7}\gamma^{8}\cdots\gamma^{12}
=\left(\ba{cc} 1&0\\0&-1\ea\right)\,,
\label{gamma7}
\ee
the six dimensional gamma matrices  are in the block diagonal form,
\be
\gamma^{a}=\left(\begin{array}{cc}0&\rho^{a}\\ \bar{\rho}^{a}&0\end{array}\right)\,,
\ee
satisfying the hermiticity conditions,
\be
\bar{\rho}^{a}=(\rho^{a})^{\dagger}\,.
\label{herm}
\ee

We can further set all the $4\times 4$ matrices,
$\rho^{a},\,\bar{\rho}^{a}$ to
be anti-symmetric~\cite{Park:1998nra}
\begin{equation}
\begin{array}{ll}
({\rho}^{a})_{\dalpha\dbeta}=-({\rho}^{a})_{\dbeta\dalpha}\,,~~~&~~~
(\bar{\rho}^{a})^{\dalpha\dbeta}=
-\half\epsilon^{\dalpha\dbeta\dgamma\ddelta}(\rho^{a})_{\dgamma\ddelta}\,,
%%%
%%-(\bar{\rho}^{a})^{\dbeta\dalpha}\,.
%%%
\end{array}\label{anti-sym}
\end{equation}
which makes the relation, ${\mbox{su}(4)\equiv\mbox{so}(6)}$,  manifest. That is,
the indices, $\dalpha,\dbeta=1,2,3,4$,
denote the fundamental representation of $\mbox{su}(4)$. \\

Note that precisely the same equations as (\ref{antisym})-(\ref{rbr})
hold for the $\mbox{so}(6)$ gamma matrices,  $\{\rho^{a},\,\bar{\rho}^{b}\}$
after replacing $\mu,\nu$, $\alpha,\beta$ by
$a,b$, $\dalpha,\dbeta$, etc.

\subsection{Ten dimensions again}
Using the four and six dimensional gamma matrices above, we write the ten dimensional gamma matrices,
\begin{equation}
\begin{array}{ll}
{{\Gamma}}^{m}=\hat{\gamma}^{m}\otimes\gamma^{(7)}&~~~\mbox{for~}~m=0,1,2,3\\
{}&{}\\
{{\Gamma}}^{a}\,=\,1\,\otimes\gamma^{a}&~~~\mbox{for~}~a=7,8,9,10,11,12\,.
\end{array}
\end{equation}

In the above choice of  gamma matrices, we have from (\ref{g11}), (\ref{4Dc}), (\ref{gamma7})
\be
\Gamma^{(10)}=\hat{\gamma}^{(5)}\otimes\gamma^{(7)}\,,
\ee
and
\be
\ba{ll}
\A=\hat{A}_{+}\otimes 1\,,~~~&~~~\B_{\pm}=\C_{\pm}\A\,,\\
{}&{}\\

\B_{+}=\hat{B}_{-}\otimes\left(\ba{cc}0&+1\\-1&0\ea\right)\,,~~~&~~~
\B_{-}=\hat{B}_{+}\otimes\left(\ba{cc}0&+1\\+1&0\ea\right)\,,\\
{}&{}\\
\C_{+}=\hat{C}_{-}\otimes\left(\ba{cc}0&-1\\+1&0\ea\right)\,,~~~&~~~
\C_{-}=\hat{C}_{+}\otimes\left(\ba{cc}0&+1\\+1&0\ea\right)\,.
\ea
\ee
Majorana spinor %, $\Psi^{\ast}=\B_{+}\Psi\,,$
is now of the form,
\be
\ba{ll}
\Psi=\B_{+}^{-1}\Psi^{\ast}=\left(\ba{c}\psi_{+}^{\alpha}{}_{\dalpha}\\{}\\
\psi_{-}^{\alpha\dalpha}\ea\right)\,,
~~~~~&~~~~(\psi_{+}^{\dagger})_{\alpha}{}^{\dalpha}=
(\hat{B}_{-})_{\alpha\beta}\psi_{-}^{\beta\dalpha}\,,
\ea
\ee
where $\alpha$ is the $\so(1,3)$ spinor index and $\pm$ denote the $\so(6)$ chirality. \\

Further to have 10 dimensional Majorana-Weyl spinor, imposing the chirality condition, $\Gamma^{(10)}\Psi=\Psi$,    we also have
\be
\hat{\gamma}^{(5)}\psi_{\pm}=\pm\psi_{\pm}\,.
\ee

For the later convenience, we define $\psi_{\alpha\dalpha}$, $\bar{\psi}^{\alpha\dalpha}$ by
\be
\ba{ll}
\psi_{\alpha\dalpha}=i(\hat{C}_{+})_{\alpha\beta}\psi_{+}^{\beta}{}_{\dalpha}\,,~~~~&~~~~
\bar{\psi}^{\alpha\dalpha}=\psi_{-}^{\alpha\dalpha}\,.
\ea
\label{12later}
\ee
The Majorana condition is equivalent to
\be
\ba{ll}
\bar{\psi}^{\alpha\dalpha}=A^{\alpha}{}_{\beta}(\psi^{\dagger})^{\beta\dalpha}\,,~~~~&~~~~
A=i\hat{A}_{-}=A^{\dagger}=A^{-1}\,.
\ea
\label{10DM}
\ee

%%%%%%%%%%%%%%%%%%%%%%%%%%%%%%%%%%%%%%%%%%%%%%%%%%%%%%%%%%%%%%%%%%%%%%%%%%%%%%%%%%%%%%%%%%%%%%%%
%%%%%%%%%%%%%%%%%%%%%%%%%%%%%
\subsection{Twelve dimensions}
In order to make the $\mbox{SO}(2,4)\times\mbox{SO}(6)$ isometry of
$AdS_{5}\times S^{5}$ geometry manifest, it is convenient to employ
the twelve dimensional gamma matrices of spacetime signature,
$(--++++++++++)$, and write them in terms of two sets of
six dimensional gamma matrices, $\{\gamma^{\mu}\}$, $\{\gamma^{a}\}$, which  we reviewed above,
\begin{equation}
\begin{array}{ll}
{\bf{\Gamma}}^{\mu}=\gamma^{\mu}\otimes\gamma^{(7)}&~~~\mbox{for~}~\mu=1,2,3,4,5,6\\
{}&{}\\
{\bf{\Gamma}}^{a}\,=\,1\,\otimes\gamma^{a}&~~~\mbox{for~}~a=7,8,9,10,11,12\,.
\end{array}
\end{equation}
%%%
%%In particular,
%%\be
%%\ba{ll}
%%{\bf\Gamma}^{1}=U\left(\ba{cc}0&+1\\-1&0\ea\right)\otimes\gamma^{(7)}\,,~~~~&~~~
%%{\bf\Gamma}^{m+2}=U\left(\ba{cc}0&\hat{\gamma}^{m}\\
%%(\hat{\gamma}^{m})^{T}&0\ea\right)\otimes\gamma^{(7)}\,,\\
%%{}&{}\\
%%{\bf\Gamma}^{6}=U\left(\ba{cc}0&\hat{\gamma}^{(5)}\\
%%(\hat{\gamma}^{(5)})^{T}&0\ea\right)\otimes\gamma^{(7)}\,,~~~~&~~~
%%U=\left(\ba{cc}\hat{C}_{+}&~0\\0&~~\hat{C}_{+}^{-1}\ea\right)\,.
%%\ea
%%\ee
%%%%

In the above choice of  gamma matrices,
the twelve dimensional charge conjugation matrices,
${\CC}_{\pm}$, are  given by
\begin{equation}
\begin{array}{ll}
\pm({\bf{\Gamma}}^{\bf M}){}^{T}={\CC}_{\pm}{\bf{\Gamma}}^{\bf M}{\CC}_{\pm}^{-1}\,,~~{\bf M}=1,2,\cdots,12,
~~&~~~{\CC}_{\pm}=\left(\begin{array}{cc}0&1\\\pm 1&0\end{array}\right)\otimes\left(\begin{array}{cc}0&1\\\mp 1&0\end{array}\right)\,,
\end{array}
\end{equation}
while the complex conjugate matrices, $\AA_{\pm}$, read
\begin{equation}
\ba{ll}
{{\AA}_{\pm}=\left(\begin{array}{cc}A^{t}&0\\  0&\mp A\end{array}\right)\otimes \left(\begin{array}{cc}1&0\\ 0&\pm 1\end{array}\right)\,,}~~~&~~A=-i\bar{\rho}_{12}=-i\hat{\gamma}^{0}=i\hat{A}_{-}=
A^{\dagger}=A^{-1}\,,
\ea\label{Apm}
\ee
satisfying
\be
\pm({\bf{\Gamma}}^{\bf M}){}^{\dagger}={\AA}_{\pm}{\bf{\Gamma}}^{\bf M}{\AA}_{\pm}^{-1}\,.
\ee
In particular, for $\mu=1,2,\cdots,6$, we have
\be
\ba{ll}
(\rho^{\mu})^{\dagger}=-A\bar{\rho}^{\mu}A^{t}=\bar{\rho}_{\mu}\,,~~~&~~~
(\bar{\rho}^{\mu})^{\dagger}=-A^{t}\rho^{\mu}A={\rho}_{\mu}\,.
\ea
\ee

Now if we define the twelve dimensional chirality operator as
\be
{\bf{\Gamma}^{(13)}}\equiv\gamma^{(7)}\otimes\gamma^{(7)}\,,
\ee
then
\begin{equation}
\ba{lll}
\{{\bf{\Gamma}}^{(13)},\,{\bf{\Gamma}}^{\bf M}\}=0\,,~~&~~\CC_{-}=\CC_{+}{\bf{\Gamma}}^{(13)}\,,~~&~~ \AA_{-}=\AA_{+}{\bf{\Gamma}}^{(13)}\,.
\ea
\ee
In 2+10 dimensions it is possible to impose the Majorana-Weyl condition on spinors
to have  sixteen independent complex components which coincides with the number of supercharges in the $AdS_{5}\times S^{5}$ superalgebra, ${\mbox{su}(2,2|4)}$.  Up to the redefinition of the spinor by a phase factor, there are essentially two choices for the Majorana-Weyl condition depending on the chirality,
\be
\ba{lll}
{\bf \Psi}=\pm{\bf{\Gamma}}^{(13)}{\bf\Psi}\,,~~&\mbox{and}&~~
\bf{\bar{\Psi}}={\bf\Psi}^{\dagger}\AA_{+}={\bf\Psi}^{T}\CC_{+}\,.
\ea
\label{MW}
\ee
Our  choice will be the plus sign so that  the $2+10$ dimensional Weyl spinor carries the same chiral indices for $\su(2,2)$ and $\su(4)$, i.e. ${\bf\Psi}=(\psi_{\alpha\dalpha}\,,~ \bar{\psi}^{\alpha\dalpha})^{T}$, while the Majorana condition relates them as $\bar{\psi}^{\alpha\dalpha}=A^{\alpha}{}_{\beta}(\psi^{\dagger})^{\beta\dalpha}$ which is identical to (\ref{10DM}). Hence, the Majorana-Weyl spinor in $2+10$ dimenisons can be identified as the Majorana spinor in $1+9$ dimensions.

\newpage
%%%%%%%%%%%%%%%%%%%%%%%%%%%%%%%%%%%%%%%%%%%%%%%%%%%%%%%%%%%%%%%%%%%%%%%%%%%%%%%%%%%%%%%%%%%%%%%%
%%%%%%%%%%%%%%%%%%%%%%%%%%%%%%%%%%%%%%%%%%%%%%%%%%%%%%%%%%%%%%%%%%%%%%%%%%%%%%%%%%%%%%%%%%%%%%%%%
%%%%%%%%%%%%%%%%%%%%%%%%%%%%%%%%%%%%%%%%%%%%%%%%%%%%%%%%%%%%%%%%%%%%%%%%%%%%%%%%%%%%%%%%%%%%%%%%%
%%%%%%%%%%%%%%%%%%%%%%%%%%%%%%%%%%%%%%%%%%%%%%%%%%%%%%%%%%%%%%%%%%%%%%%%%%%%%%%%%%%%%%%%%%%%%%%%%
%%%%%%%%%%%%%%%%%%%%%%%%%%%%%%%%%%%%%%%%%%%%%%%%%%%%%%%%%%%%%%%%%%%%%%%%%%%%%%%%%%%%

\section{Looking for the general odd symmetry\label{ProofSec}}
With a Majorana-Weyl spinor, $\E$, $\DP$, which may depend on $x^{M}$, we focus on the following transformations,
\be
\ba{ll}
\delta A_{M}=i\bar{\Psi}\Gamma_{M}\E=-i\bar{\E}\Gamma_{M}\Psi\,,~~&~~\delta\Psi=\half F_{MN}\Gamma^{MN}\E+\DP\,,
\ea
\ee
so that
\be
\delta\bar{\Psi}=-\half\bar{\E}F_{MN}\Gamma^{MN}+\overline{\DP}\,.
\ee
Note that $\DP$ is Lie algebra valued, while $\E$ is not.\\

From
\be
\Psi^{p\alpha}\Psi^{q\beta}\Psi^{r\gamma}\tr(T_{p}T_{q}T_{r})=
\Psi^{p\gamma}\Psi^{q\alpha}\Psi^{r\beta}\tr(T_{p}T_{q}T_{r})=
\Psi^{p\beta}\Psi^{q\gamma}\Psi^{r\alpha}\tr(T_{p}T_{q}T_{r})\,,
\ee
and the identity (\ref{crucial}), we note that the second term in (\ref{infL}) vanishes
\be
\tr\left(\bar{\Psi}\Gamma^{M}\Psi\,\bar{\Psi}\Gamma_{M}\E\right)=0\,.
\ee
\\
We also get, using the Bianchi identity (\ref{Bianchi}),
\be
\ba{ll}
\bar{\Psi}\Gamma^{M}D_{M}\delta\Psi&=\half D_{L}F_{MN}\bar{\Psi}(\Gamma^{LMN}+2\eta^{LM}\Gamma^{N})\E
+\half\bar{\Psi}\Gamma^{L}\Gamma^{MN}\partial_{L}\E F_{MN}+\bar{\Psi}\Gamma^{L}D_{L}\DP\\
{}&{}\\
{}&=-iD_{M}F^{MN}\delta A_{N}+\half\bar{\Psi}\Gamma^{L}\Gamma^{MN}\partial_{L}\E F_{MN}+
\bar{\Psi}\Gamma^{L}D_{L}\DP\,.
\ea
\ee
\\

Thus, semi-finally,  we obtain
\be
%%\ba{ll}
\delta\L=-i\,\tr\left[\half F_{MN}\bar{\Psi}\Gamma^{L}\Gamma^{MN}\partial_{L}\E+
\bar{\Psi}\Gamma^{L}D_{L}\DP\right]
+\partial_{N}\tr\left[F^{MN}\delta A_{M}+i\half\bar{\Psi}\Gamma^{N}\delta\Psi\right]\,.
%%
%%{}&{}\\
%%{}&=-i\half\tr\left[F_{MN}\bar{\Psi}(
%%\Gamma^{MN}\Gamma^{L}\partial_{L}-4\Gamma^{[M}\partial^{N]})\E\right]
%%+\partial_{N}\tr\left[F^{MN}\delta A_{M}+i\half\bar{\Psi}\Gamma^{N}\delta\Psi\right]
%%\ea
%%%
\label{infL2}
\ee
{}\\
We first note that constant $\E$, and constant $\DP$ which is central in the Lie algebra lead to the ordinary  and kinetic supersymmetries
\be
\ba{lll}
\E,~\DP~~:~\mbox{constant}~~&~~\mbox{and}~~&~~\DP\propto \I_N\,.
\ea
\label{OK}
\ee

Henceforth, keeping {\it the dimensional reduction either to  Minkowskian $d$-dimensions, $0\leq m\leq {d-1}$, $d\leq a\leq 9$, or Euclidean $d$-dimensions, $1\leq m\leq d$, $a=0$, $d+1\leq a\leq 9$,}  we set  $A_{a}=\Phi_{a}$,  ``\,$\partial_{a}\equiv 0$\,'', and look for some possibilities of more general  symmetries. \\
\\
Since
\be
F_{MN}\Gamma^{L}\Gamma^{MN}\partial_{L}\E=
\left(F_{mn}\Gamma^{l}\Gamma^{mn}+
2D_{m}\Phi_{b}\Gamma^{l}\Gamma^{m b}+
D_{a}\Phi_{b}\Gamma^{l}\Gamma^{ab}\right)\partial_{l}\E\,,
\label{c0}
\ee
we first require
\be
\Gamma^{l}\Gamma^{mn}\partial_{l}\E=0\,,
\label{c1}
\ee
or equivalently
\be
\Gamma^{mn}\Gamma^{l}\partial_{l}\E=
2\Gamma^{m}\partial^{n}\E-2\Gamma^{n}\partial^{m}\E\,.
\label{c2}
\ee
It follows after multiplying $\Gamma_{nm}$ without $m,n$ summing,
\be
\Gamma^{l}\partial_{l}\E=
2\Gamma^{m}\partial_{m}\E+2\Gamma^{n}\partial_{n}\E~~~:~~\mbox{no\,~sum\,~for\,~}
m\neq n\,.
\label{c3}
\ee
Eqs.(\ref{c1}),\,(\ref{c2}),\,(\ref{c3}) are trivial when $d=0,1$.  For $d\geq 2$, summing over $m\neq n$ in (\ref{c3}) we get
\be
(d-1)(d-4)\Gamma^{l}\partial_{l}\E=0\,.
\ee
Hence, for $d=2,3$, $d\geq 5$,
\be
\Gamma^{m}\partial_{m}\E=-\Gamma^{n}\partial_{n}\E~~~:~~\mbox{no\,~sum\,~and\,~}
m\neq n\,.
\label{c4}
\ee
\begin{itemize}
\item For $d=3$, $d\geq 5$, we easily conclude $\partial_{m}\E=0$, i.e.  constant parameter, $\E$.

\item When $d=2$, we get
\be
\partial_{m}\E=-\Gamma_{mn}\partial^{n}\E~~~:~~\mbox{for~}d=2\,,
\label{d2E}
\ee
so that
\be
\partial^{m}\partial_{m}\E=0\,.
\label{d2E2}
\ee
Let $\sigma\neq\tau$ be the two different spacetime indices in $d=2$ case.  Eq.(\ref{c1}) is   simply  equivalent to
\be
(\partial_{\sigma}+\Gamma_{\sigma}{}^{\tau}\partial_{\tau})\E=0\,.
\ee
This can be solved easily in the diagonal basis of $\Gamma_{\sigma}{}^{\tau}$. In the Minkowskian two-dimensions, as $\Gamma_{0}{}^{1}$ is hermitian, the solution is given by the left and right modes, ${\sigma\pm\tau}$. On the other hand, in the Euclidean  two-dimensions, $\Gamma_{1}{}^{2}$ is anti-hermitian and the solution involves  holomorphic functions, $\sigma\pm i\tau$.

\item For $d=4$ we have for any  $m$,
\be
\ba{ll}
\partial_{m}\E=\Gamma_{m}\xi_{-}\,,~~~&~~~\xi_{-}=\textstyle{\frac{1}{4}}
\Gamma^{l}\partial_{l}\E\,.
\ea
\ee
From $\partial_{[m}\partial_{n]}\E=0$ we get  an essenitally same relation as  (\ref{c4}),
\be
\Gamma^{m}\partial_{m}\xi_{-}=-\Gamma^{n}\partial_{n}\xi_{-}~~~:~~\mbox{no\,~sum\,~and\,~}
m\neq n\,.
\label{c5}
\ee
Hence, $\xi_{-}$ is a  constant spinor,  and
\be
\E=x^{m}\Gamma_{m}\xi_{-}+\xi_{+}\,,
\label{4dE}
\ee
where $\xi_{+},\xi_{-}$ are constant Majorana-Weyl spinors of the opposite chiralities, corresponding to the ordinary supersymmetry and special superconformal symmetry, respectively.
\end{itemize}

Provided the above solutions for (\ref{c1}),  we are  ready for the full analysis.

\begin{enumerate}
\item  When $d=0$ : IKKT matrix model. \\
Eq.(\ref{c0}) becomes trivial, and we natually require
\be
\Gamma^{a}[\Phi_{a},\DP]=0\,.
\label{GPx}
\ee
We need to  find the algebraic solution for $\DP$ in terms of the  Lie algebra valued fields, $\Phi_{a}$, $d\leq a\leq 9$. Clearly, the kinetic supersymmetry transformation, i.e.  $\DP\propto\I_N$, satisfies the above equation. In fact, we can show that this is the most general solution. \\
{}\newline
{\it Proof}\\
We consider the special case, $\Phi_{a}=0$, $d\leq a\leq 7$. Eq.(\ref{GPx}) gives
\be
[\Phi_{8},\Gamma^{8}\DP]+[\Phi_{9},\Gamma^{9}\DP]=0\,.
\ee
Multiplying $\Phi_{8}$ and taking the  $\mbox{u}(N)$ trace we get
\be
\tr\left([\Phi_{8},\Phi_{9}]\DP\right)=0\,.
\ee
Since the commutator, $[\Phi_{8},\Phi_{9}]$, can be arbitrary except $\I_N$, we conclude that $\DP\propto\I_N$. This completes our proof.\\
\newline
Therefore, when $d=0$, $\E$ and $\DP$ are simply  constant Majorana-Weyl spinors  corresponding to the ordinary and the kinetic supersymmetries.

\item When $d=1$ : BFSS matrix model. \\
Eq.(\ref{c1}) is trivial, and with the coordinate, $\tau$ for $d=1$, From Eq.(\ref{infL2}) we require
\be
\ba{ll}
0&=\half F_{MN}\Gamma^{L}\Gamma^{MN}\partial_{L}\E+\Gamma^{L}D_{L}\DP\\
{}&{}\\
{}&=\Gamma^{\tau}D_{\tau}(\DP+\Phi_{a}\Gamma^{\tau a}\partial_{\tau}\E)
+\Gamma^{b}D_{b}(\DP-\half\Phi_{a}\Gamma^{\tau a}\partial_{\tau}\E)
-\Phi_{a}\Gamma^{a}\partial^{\tau}\partial_{\tau}\E\,.
\ea
\ee
The only possible algebraic solutions are (\ref{OK}) corresponding to the ordinary and the kinetic supersymmetries.

\item When $d=2$.\\
From (\ref{infL2}) we require,  using   (\ref{d2E}), (\ref{d2E2}),
\be
\ba{ll}
0&=\half F_{MN}\Gamma^{L}\Gamma^{MN}\partial_{L}\E+\Gamma^{L}D_{L}\DP\\
{}&{}\\
{}&=\Gamma^{m}D_{m}\DP+\Gamma^{a}D_{a}\DP+
2(D^{\tau}\Phi_{a}-D^{\sigma}\Phi_{a}\Gamma_{\sigma}{}^{\tau})\Gamma^{a}\partial_{\tau}\E\,.
\ea
\ee
We conclude again that the only possible algebraic solutions are (\ref{OK}) corresponding to the ordinary and the kinetic supersymmetries.

\item When $d=3$, $d\geq 5$.\\
Since $\E$ is constant, the only possible algebraic solutions are (\ref{OK}) corresponding to the ordinary and the kinetic supersymmetries.

\item When $d=4$.\\
From (\ref{infL2}) we require,  using   (\ref{4dE}),
\be
\ba{ll}
0&=\half F_{MN}\Gamma^{L}\Gamma^{MN}\partial_{L}\E+\Gamma^{L}D_{L}\DP\\
{}&{}\\
{}&=\Gamma^{L}D_{L}(\DP+2\Phi_{a}\Gamma^{a}\xi_{-})\,.
\ea
\ee
Thus the algebraic solution reads
\be
\DP+2\Phi_{a}\Gamma^{a}\xi_{-}\propto \I_N\,.
\ee
\end{enumerate}

~\\
~\\

\section*{Acknowledgments} 
This work is  supported by Basic Science Research Program through the National Research Foundation of Korea  Grants, NRF-2016R1D1A1B01015196 and NRF-2020R1A6A1A03047877 (Center for Quantum Space Time).

\newpage
\bibliographystyle{unsrt}
\bibliography{reference}

\end{document}